\documentclass[letter]{aa}  
\usepackage{natbib}
\usepackage{CJK}

\usepackage{graphicx}
\usepackage{txfonts}
\usepackage{xcolor}
\usepackage{multirow}
\def\h2{\ion{H}{II}}

\usepackage{lipsum}
\usepackage{tikz,xcolor}
\usepackage[colorlinks=true, allcolors=blue]{hyperref}
\definecolor{lime}{HTML}{A6CE39}
\DeclareRobustCommand{\orcidicon}{
        \begin{tikzpicture}
        \draw[lime, fill=lime] (0,0) 
        circle [radius=0.16] 
        node[white] {{\fontfamily{qag}\selectfont \tiny ID}};
        \draw[white, fill=white] (-0.0625,0.095) 
        circle [radius=0.007];
        \end{tikzpicture}
        \hspace{-2mm}
}
\foreach \x in {A, ..., Z}{\expandafter\xdef\csname orcid\x\endcsname{\noexpand\href{https://orcid.org/\csname orcidauthor\x\endcsname}
                        {\noexpand\orcidicon}}
}

\begin{document}

   \title{Discovery of non-metastable ammonia masers in Sagittarius B2}


   \author{{\protect\begin{CJK*}{UTF8}{gkai}Y. T. Yan (闫耀庭)\protect\end{CJK*}} \inst{\ref{inst.mpifr}}\fnmsep\thanks{Member of the International Max Planck Research School (IMPRS) for Astronomy and Astrophysics at the universities of Bonn and Cologne.}\orcidB{}
         \and C. Henkel\inst{\ref{inst.mpifr},\ref{inst.KingAbdulazizU},\ref{inst.xao}}\orcidA{}
         \and K. M. Menten\inst{\ref{inst.mpifr}}\orcidC{}
         \and {\protect\begin{CJK*}{UTF8}{gkai}Y. Gong (龚\protect\end{CJK*}\protect\begin{CJK*}{UTF8}{bkai}龑\protect\end{CJK*})} \inst{\ref{inst.mpifr}}\orcidD{}
         \and H. Nguyen\inst{\ref{inst.mpifr}}\fnmsep$^\star$\orcidK{}
         \and J. Ott\inst{\ref{inst.nrao}}\orcidE{}
         \and A. Ginsburg\inst{\ref{inst.uf}}\orcidF{}
         \and T. L. Wilson\inst{\ref{inst.mpifr}}
         \and A. Brunthaler\inst{\ref{inst.mpifr}}\orcidH{}
         \and A. Belloche\inst{\ref{inst.mpifr}}\orcidI{}
         \and {\protect\begin{CJK*}{UTF8}{gkai}J. S. Zhang (张江水)\protect\end{CJK*}}\inst{\ref{inst.gzhu}}\orcidJ{}
         \and N. Budaiev\inst{\ref{inst.uf}}
         \and D. Jeff\inst{\ref{inst.uf}}
         }

   \institute{
\label{inst.mpifr}Max-Planck-Institut f\"{u}r Radioastronomie, Auf dem H\"{u}gel 69, 53121 Bonn, Germany\\  \email{yyan@mpifr-bonn.mpg.de, astrotingyan@gmail.com}
\and\label{inst.KingAbdulazizU}Astronomy Department, Faculty of Science, King Abdulaziz University, P.~O.~Box 80203, Jeddah 21589, Saudi Arabia
\and\label{inst.xao}Xinjiang Astronomical Observatory, Chinese Academy of Sciences, 830011 Urumqi, PR China
\and\label{inst.nrao}National Radio Astronomy Observatory, 520 Edgemont Road, Charlottesville, VA 22903-2475, USA
\and\label{inst.uf}Department of Astronomy, University of Florida, PO Box 112055, USA
\and\label{inst.gzhu}Center for Astrophysics, Guangzhou University, 510006 Guangzhou, People's Republic of China
             }

   \date{Received 20 September; accepted 23 September}

 
  \abstract
  {We report the discovery of widespread maser emission in non-metastable inversion transitions of NH$_3$ toward various parts of the Sagittarius B2 molecular cloud and star-forming region complex. We detect masers in the $J,K = $ (6,3), (7,4), (8,5), (9,6), and (10,7) transitions toward Sgr B2(M) and Sgr B2(N), an NH$_3$ (6,3) maser in Sgr B2(NS), and NH$_3$ (7,4), (9,6), and (10,7) masers in Sgr B2(S). With the high angular resolution data of the Karl G. Jansky Very Large Array (JVLA) in the A-configuration, we identify 18 maser spots. Nine maser spots arise from Sgr B2(N), one from Sgr B2(NS), five from Sgr B2(M), and three in Sgr B2(S). Compared to our Effelsberg single-dish data, the JVLA data indicate no missing flux. The detected maser spots are not resolved by our JVLA observations. Lower limits to the brightness temperature are $>$3000~K and reach up to several 10$^5$~K, manifesting the lines' maser nature. In view of the masers' velocity differences with respect to adjacent hot molecular cores and/or UCH{\scriptsize II} regions, it is argued that all the measured ammonia maser lines may be associated with shocks caused either by outflows or by the expansion of UCH{\scriptsize II} regions. Overall, Sgr B2 is unique in that it allows us to measure many NH$_3$ masers simultaneously, which may be essential in order to elucidate their thus far poorly understood origin and excitation.}


   \maketitle
%

\section{Introduction}
\label{introduction}

Since their discovery in the $(J,K)$~=~(3,3) metastable ($J=K$) line toward the high-mass star-forming region (HMSFR) W33 \citep{1982A&A...110L..20W}, sources emitting maser emission in the ammomia molecule (NH$_3$) have attracted much attention. To date, metastable ammonia maser lines have been detected in 22 HMSFRs \citep[see][and references therein]{2022A&A...659A...5Y}, while non-metastable ($J>K$) ammonia masers have only been found in ten sources. The NH$_3$ (6,3), (7,4), (8,5), and (9,6) maser lines, which will be discussed below, arise from energy levels of 551~K, 713~K, 892~K, and 1089~K above the ground state, respectively. These four maser transitions have  only been detected, respectively, in three (NGC7538, W51, and Sgr B2(N)), in two (W51 and Sgr B2(N)), in two (W51 and Sgr B2(N)), and in seven (W51, NGC7538, W49, DR21 (OH), Sgr B2(N), Cep A and G34.26$+$0.15) HMSFRs \citep{1986ApJ...300L..79M,2013A&A...549A..90H,2020ApJ...898..157M,2022A&A...659A...5Y}. The NH$_3$ (10,7) line, also observed by us, connects states as high as 1303 K above the ground state and has so far only been classified as a maser in W51 \citep{2013A&A...549A..90H}. Among all the abovementioned regions of massive star formation, Sgr B2 hosts a particularly high number of active sites of star formation. 

\begin{figure*}[h]
\center
    \includegraphics[width=450pt]{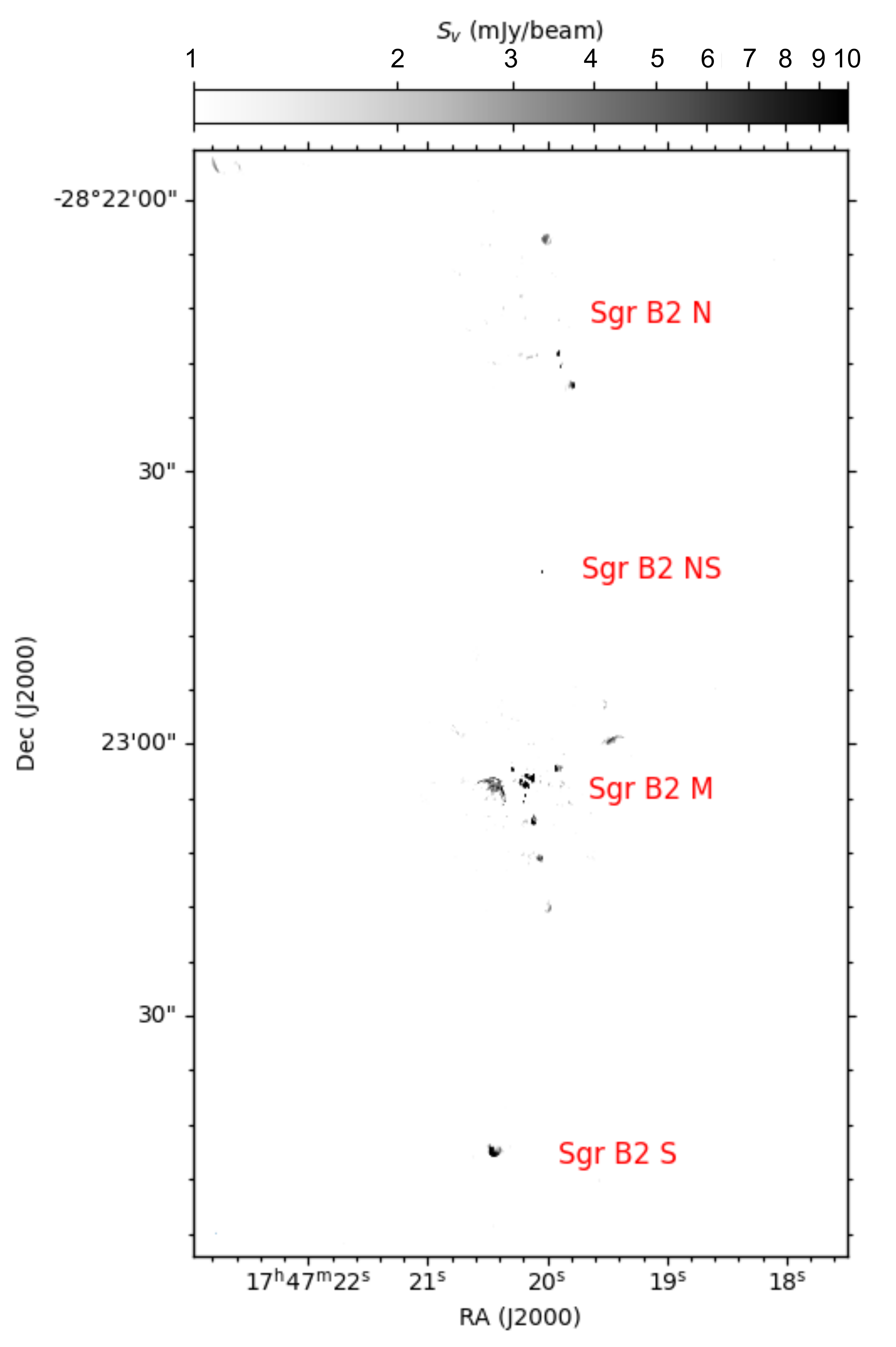}
  \caption{\label{1.6cm-continuum}JVLA 1.6-cm continuum map of Sgr B2, shown in gray. The synthesized beam is $0\farcs22 \times 0\farcs08$, P.A.~=~$-10\fdg61$.}
\end{figure*}

Sgr B2 is located at a projected distance of $\sim$100 pc from Sgr A$^ {\rm *}$ \citep{2009ApJ...705.1548R}, the compact radio source associated with the supermassive black hole in the Galactic center at a distance of 8.178$\pm$0.013$\rm _{stat.}\pm$ 0.022$\rm _{sys.}$ kpc \citep{2019A&A...625L..10G}. This region is normally divided into three high-mass star-forming cores: Sgr B2(N), Sgr B2(M), and Sgr B2(S) (see our Fig.~\ref{1.6cm-continuum} for the locations of these). In total, Sgr B2 contains more than 50 \h2 regions, most of which are ultracompact \h2 regions (UC\h2) with diameters smaller than 0.1 pc \citep{1995ApJ...449..663G,1998ApJ...500..847D,2014ApJ...781L..36D,2015ApJ...815..123D,2008ApJS..174..481L,2018ApJ...853..171G,2019A&A...630A..73M,2022arXiv220807796M,2021A&A...651A..88N}. In Sgr B2 (N) and (M), several deeply embedded high-mass young stellar objects are surrounded by dense, hot molecular cores with an exceedingly rich chemistry that gives rise to a plethora of lines from numerous complex organic molecules \citep[e.g.,][]{2008A&A...482..179B,2013A&A...559A..47B,2022A&A...662A.110B}. Detected molecular maser species are OH \citep{1990ApJ...351..538G,2013MNRAS.431.1180C,2016ApJS..227...10C}, H$_2$O \citep{1988ApJ...330..809R,2004ApJS..155..577M,2014MNRAS.442.2240W}, 
SiO \citep{1992PASJ...44..373M,2009ApJ...691..332Z}, H$_2$CO \citep{1994ApJ...434..237M,2007ApJ...654..971H,2019ApJS..244...35L}, CH$_3$OH class I \citep{1997ApJ...474..346M,2016ApJS..227...10C}, and class II \citep{1996MNRAS.283..606C,2019MNRAS.482.5349R,2019ApJS..244...35L},  and NH$_3$ \citep{1999ApJ...519..667M,2018ApJ...869L..14M,2020ApJ...898..157M}. In the case of NH$_3$, an ammonia maser in the metastable (3,3) transition was only detected in the southern part of Sgr B2(S) \citep{1999ApJ...519..667M}. Another metastable ammonia maser, in the (2,2) line, was found toward SgrB2(M) \citep{2018ApJ...869L..14M}. Recently, 18 ammonia non-metastable maser lines at frequencies of 13.0 -- 24.0 GHz were detected toward Sgr B2(N) with the Shanghai 65-meter Tianma radio telescope with characteristic beam sizes of 54 $\times$ 18.5/$\nu$(GHz) arcseconds \citep{2020ApJ...898..157M}. 

In this letter, we report the discovery of NH$_3$ (6,3), (7,4), (8,5), (9,6), and (10,7) masers in Sgr B2(M) and Sgr B2(N), an NH$_3$ (6,3) maser in Sgr B2(NS), as well as NH$_3$ (7,4), (9,6), and (10,7) masers in Sgr B2(S). All of these increase the number of (6,3), (7,4), (8,5), (9,6), and (10,7) maser detections in our Galaxy, respectively, from three to six, two to four, two to three, seven to nine, and one to four. Observations with the Effelsberg 100-meter telescope and the Karl G. Jansky Very Large Array (JVLA) are presented in Sect. \ref{observations}. Results are described in Sect. \ref{results}. A comparison of the positions of the different ammonia masers with those of other relevant tracers of the interstellar medium is presented in Sect. \ref{disscussion}. Our main results are summarized in Sect. \ref{summary}.

\section{Observations and data reduction}
\label{observations}

\subsection{Effelsberg observations and data reduction}
\label{EffelsbergObservations}

The NH$_3$ (9,6) and (10,7) lines were observed toward Sgr B2 with the 100-meter Effelsberg telescope\footnote{Based on observations with the 100-meter telescope of the MPIfR (Max-Planck-Institut f\"{u}r Radioastronomie) at Effelsberg.} at 12 epochs in January 2020, February and August 2021, as well as in March, May, June, July, and August 2022. The observations were performed in position switching mode. The off position was 30$\arcmin$ in azimuth away from the source. An S14mm double beam secondary focus receiver was employed. The half power beam width (HPBW) is 49 $\times$ 18.5/$\nu$(GHz) arcseconds, that is 49$\arcsec$ at 18.5 GHz, the frequency of the NH$_3$ (9,6) line. Before August 2021, the spectrometer covered 2\,GHz with a channel width of 38.1 kHz, corresponding to $\sim$0.62~km~s$^{-1}$ at 18.5 GHz. From August 2021, a high spectral resolution backend with 65536 channels and a bandwidth of 300 MHz was employed, providing a channel width of 0.07~km s$^{-1}$ at 18.5 GHz. Pointing was checked every hour using NGC~7027. Focus calibrations were done at the beginning of the observations, and during sunset and sunrise, toward NGC~7027. The calibrator was measured between elevations of 30 and 56 degrees. The elevation on target was about 10 degrees, requiring only minimal ($\leq$~2\%) elevation-dependent flux density corrections. The system temperatures were 140--220 K in a main-beam brightness temperature, $T_{\rm MB}$, scale. The flux density was calibrated assuming a $T_{\rm MB}/S$ ratio of 1.95 K/Jy, derived from continuum cross scans of NGC 7027 (its flux density was adopted from \citealt{1994A&A...284..331O}). Calibration uncertainties were estimated to be $\sim10$\%. 

We used the GILDAS/CLASS\footnote{https://www.iram.fr/IRAMFR/GILDAS/} package \citep{2005sf2a.conf..721P} to reduce the spectral line data. A first-order polynomial was subtracted from each spectrum for baseline removal. 

\subsection{JVLA observations and data reduction}
\label{JVLAObservations}

Observations of six NH$_3$ lines, the (5,1), (6,3), (7,4), (8,5), (9,6), and (10,7) transitions (Table~\ref{observedlines}), toward Sgr B2 were made on 5\ March 2022 with the JVLA of the National Radio Astronomy Observatory\footnote{The National Radio Astronomy Observatory is a facility of the National Science Foundation, operated under a cooperative agreement by Associated Universities, Inc.} (NRAO) in the A-configuration (project ID:  22A-106, PI: Yaoting Yan). Eight-bit samplers were used to perform the observations. For the NH$_3$ (9,6) and (10,7) line observations, we used two subbands with the eight-bit samplers covering a bandwidth of 16 MHz with full polarization, eight recirculations, and four baseline board pairs (BLBPs) to provide a velocity range of 260~km~s$^{-1}$ with a channel spacing of ~0.13~km~s$^{-1}$. Four additional subbands of bandwidth 16 MHz were used to cover the NH$_3$ (5,1), (6,3), (7,4), and (8,5) lines. The remaining ten subbands of the eight-bit sampler with a bandwidth of 128 MHz were used to measure the continuum emission between 18 and 20 GHz. The primary beam of the JVLA antennas is $150^{\prime \prime}$ (FWHM) at 18.5 GHz, covering all prominent star-forming regions in Sgr B2 simultaneously. 3C~286 was used as a calibrator for pointing, flux density, bandpass, and polarization \citep{2013ApJS..204...19P}. J1745-2900 served as our gain calibrator during the observations. The on-source time was 30 minutes toward Sgr B2.

A total of 27 antennas were employed for the observations. Data from two antennas were lost due to technical issues. The data from the remaining 25 antennas were reduced through the Common Astronomy Software Applications package (CASA\footnote{https://casa.nrao.edu/}; \citealt{2007ASPC..376..127M}). We calibrated the data with the JVLA CASA calibration pipeline using CASA 6.2.1. The results were obtained after flagging and recalibrating data that contained artifacts. We inspected the phase, amplitude, and bandpass variations of the calibrated visibility data to search for additional artifacts before imaging. Then, the \textit{uvcontsub} task in CASA was used to separate the calibrated visibilities into two parts: one with line-only data and the other with the line-free continuum data. The \textit{tclean} task with a cell size of 0$\farcs$02 and Briggs weighting with robust=0.5 was used to produce the images of spectral line and continuum emission. All of the images were corrected for primary beam response. The synthesized beams and the rms noises in a channel image for the observed lines are listed in Table~\ref{observedlines}. For the 1.6\,cm (18--20 GHz) continuum emission, the synthesized beam is $0\farcs22 \times 0\farcs08$ at P.A.~=~$-10\fdg61$. The typical absolute astrometric accuracy of the JVLA is $\sim$10\% of the synthesized beam\footnote{https://science.nrao.edu/facilities/vla/docs/manuals/oss/performance-/positional-accuracy}. The flux density scale calibration accuracy is estimated to be within 15\%.

The maser spots were identified in two different ways and then cross-checked. First,  we searched for masers using eyes in channel maps with a velocity spacing of 0.5~km~s$^{-1}$. Second, we used an automated source extraction code \citep[SEC;][]{murugeshan2015,nguyen2015} running in CASA 5.4 to find the maser features. Emission with signal-to-noise ratios (S/Ns) larger than six identified in this way was considered to be a real detection. A detailed description of the SEC code can be found in Sect.~3.1 of \citet{2022arXiv220710548N}.

\begin{table*}[h]
\caption{Summary of the JVLA observations.}
\centering
\begin{tabular}{ccccccccc}
\hline\hline
Transition  & $\nu$ &  $E_{\rm low}/k$ & Baseband  & Synthesized beam & Linear resolution & P.A. &  & rms \\
($J,K$) &(GHz)  & (K) &   & (arcsec) &  (au) & (deg)  & (K mJy$^{-1}$) &  (mJy beam$^{-1}$) \\
\hline
\label{observedlines}
 (5,1)    &  19.838346    &     422     &   A0/C0  & 0.198 $\times$ 0.085 & 1624 $\times$ 697 & $-$9.807 & 184 & 3.32 \\
 (6,3)    &  19.757538    &     551     &   A0/C0  & 0.203 $\times$ 0.087 & 1665 $\times$ 713 & $-$9.806 & 177 & 4.30 \\
 (7,4)    &  19.218465    &     713     &   A0/C0  & 0.209 $\times$ 0.089 & 1714 $\times$ 730 & $-$9.661 & 178 & 3.85 \\
 (8,5)    &  18.808507    &     892     &   B0/D0  & 0.214 $\times$ 0.091 & 1755 $\times$ 746 & $-$9.527 & 177 & 3.54 \\
 (9,6)    &  18.499390    &     1089    &   B0/D0  & 0.215 $\times$ 0.097 & 1763 $\times$ 795 & $-$7.385 & 171 & 3.79 \\
 (10,7)   &  18.285434    &     1303    &   B0/D0  & 0.220 $\times$ 0.094 & 1804 $\times$ 771 & $-$9.874 & 177 & 2.95 \\
\hline
\end{tabular}
\tablefoot{Columns (1) and (2): observed lines and corresponding rest frequencies, taken from \citet{2013A&A...549A..90H}. Column (3): $E_{\rm low}/k$: energy above the ground state of the lower level of a given inversion doublet; $k$ is the Boltzmann constant; $(E_{\rm up} - E_{\rm low})/k \sim$ 1.0--1.5K. Column (4): baseline board pairs' setup, see details in VLA resources page. Column (5): synthesized beam. Column (6): linear resolution at a distance of 8.2~kpc. Column (7): position angle of the synthesized beam. Column (8): conversion factor from mJy to Kelvin for each transition was calculated with 1.222~$\times$10$^3$/($\nu^2\theta_{\rm maj}\theta_{\rm min}$), where $\nu$ is the frequency in units of GHz; $\theta_{\rm maj}$ and $\theta_{\rm min}$ are the major and minor axis of the synthesized beam in units of arcseconds\footnote{https://science.nrao.edu/facilities/vla/proposing/TBconv}. Column (9): rms noise in a channel image.}
\end{table*}

\section{Results}
\label{results}

In January 2020, with the Effelsberg 100-m telescope, we observed two strong NH$_3$ (9,6) maser features with velocities of $\sim$77 km s$^{-1}$ and $\sim$84 km s$^{-1}$, and a weaker one at $\sim$72 km s$^{-1}$, toward the equatorial position $\alpha_{\rm J2000}$~=~17$^{\rm h}$47$^{\rm m}$20$\fs$8, and $\delta_{\rm J2000}$~=~$-$28$\degr$23$\arcmin$32$\farcs$1, which is offset by ($+3.^{\hspace{-1mm}\prime \prime}96$,$-26.^{\hspace{-1mm}\prime \prime}19$) from Sgr B2(M). In addition, maser emission at $\sim$77 km s$^{-1}$ was also found in the non-metastable para-NH$_3$ (10,7) transition. The NH$_3$ (9,6) and (10,7) maser spectra are shown in Fig.~\ref{spectra-2020}. In February 2021, we extended our observations to 20 positions to cover, in a fully sampled way, an area of $\sim$5.0 square arc minutes surrounding Sgr B2(M) with a spacing of 25$^{\prime \prime}$, half the beam size. (9,6) maser emission was found to be quite widespread in Sgr B2, not only residing in Sgr B2(M), but also in Sgr B2(N) and Sgr B2(S), while (10,7) masers were detected in a more limited region comprising Sgr B2(M) and Sgr B2(N). The maps of NH$_3$ (9,6) and (10,7) spectra are presented in Figs.~\ref{96spectra-2021} and \ref{107spectra-2021}.

Effelsberg monitoring observations spanning 19 months show that the NH$_3$ (9,6) maser at $V_{\rm LSR}$~=~72.5~km~s$^{-1}$ toward Sgr B2(N) became weaker from February to August in 2021 and was not detectable from March 2022 on 3$\sigma$ levels of 0.12 Jy with a 0.07~km~s$^{-1}$ channel width. A weaker (9,6) feature at $V_{\rm LSR}$~=~63.8~km~s$^{-1}$ was detected in March 2022. The (9,6) maser spectra from Sgr B2(N) are presented in Fig.~\ref{spectra-sgrb2n}. NH$_3$ (9,6) line parameters obtained by Gaussian fits are listed in Table~\ref{spectra96_fitting-sgrb2n}. An NH$_3$ (10,7) maser was detected at a different velocity of 82.0~km~s$^{-1}$ toward Sgr B2(N). Its flux density was increasing from February to August 2021 but was decreasing from March to August 2022. In Table~\ref{spectra107_fitting-sgrb2n}, NH$_3$ (10,7) line parameters obtained from Gaussian fits are presented.  

Toward Sgr B2(M), NH$_3$ (9,6) maser emission at $V_{\rm LSR}$~=~72.5~km~s$^{-1}$ became stronger between February 2021 and March 2022, then weakened until 2022 August. Higher spectral resolution data since March 2022\ show the NH$_3$ (9,6) emission to be composed of three different components. The NH$_3$ (10,7) maser in Sgr B2(M) has a velocity offset with respect to the (9,6) maser, with a velocity of $V_{\rm LSR}$~=~70.0~km~s$^{-1}$. The flux density in the Effelsberg beam remained constant within the uncertainties during the 19 months. Spectra are shown in Fig.~\ref{spectra-sgrb2m} and line parameters obtained by Gaussian fits are listed in Tables~\ref{spectra96_fitting-sgrb2m} and \ref{spectra107_fitting-sgrb2m}.

The 1.6\,cm continuum, derived from our JVLA A-configuration measurements, is shown in Fig.~\ref{1.6cm-continuum}. A total of 22 known compact \h2 regions \citep{1995ApJ...449..663G,1998ApJ...500..847D,2014ApJ...781L..36D,2015ApJ...815..123D} were detected by our observations. The locations and sizes of these sources, derived with the \textit{imfit} task in CASA, are consistent with previous results from 7 mm continuum measurements \citep{2015ApJ...815..123D}. Details are given in Table~\ref{continuum_sou}.

The JVLA has a better angular resolution compared to the Effelsberg 100-m single dish and those data reveal 18 maser spots in the NH$_3$ (6,3), (7,4), (8,5), (9,6), and (10,7) transitions. We did not find any emission in the NH$_3$ (5,1) line from Sgr B2. The 3$\sigma$ upper limit for the NH$_3$ (5,1) line is 9.96 mJy~beam$^{-1}$ (about 1800~K) for a channel width of 0.12~km~s$^{-1}$. The JVLA NH$_3$ (9,6) and (10,7) line profiles toward Sgr B2(M), extracted from an Effelsberg-beam-size region (FWHM, 49$\arcsec$), are shown in Fig.~\ref{spectra-sgrb2m}. From the similarity of the flux density obtained at Effelsberg and the JVLA, measured in March 2022, we conclude that there is no ``missing spacing'' flux density in the JVLA data, that is, emission on angular scales larger than defined by the shortest JVLA baseline. NH$_3$ (6,3) masers arise from four different locations, named 63A, 63B, 63C, and 63D. NH$_3$ (7,4), (8,5), (9,6), and (10,7) masers are detected toward three, two, four, and five spots, respectively. Positions and spectral parameters of these masers are listed in Table~\ref{NH3-positions}. The detected isolated maser spots are distributed over a $16'' \times 93''$ area in Sgr B2, corresponding to $0.6 \times 3.7$ pc. Details related to the individual sources are given below. 

\begin{figure*}[h]
\center
    \includegraphics[width=500pt]{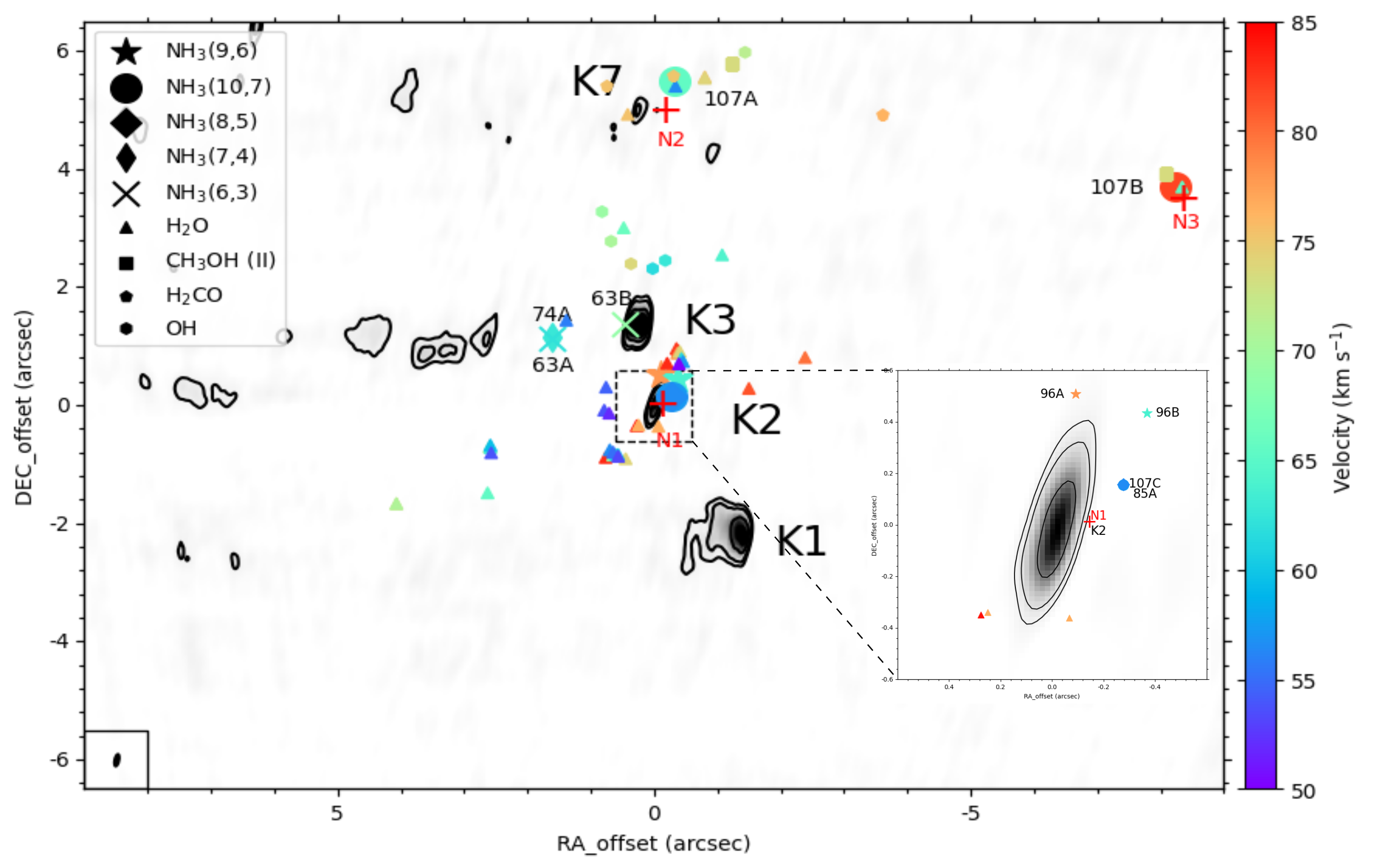}
    \caption{JVLA 1.6\,cm continuum map of Sgr B2(N), shown by the gray shaded areas and black contours with levels of 5, 10, 30, and 50~$\times$~0.2~mJy~beam$^{-1}$. The reference position is $\alpha_{\rm J2000}$~=~17$^{\rm h}$47$^{\rm m}$19$\fs$883, and $\delta_{\rm J2000}$~=~$-$28$\degr$22$\arcmin$18$\farcs$412, the peak position of the continuum source K2. The crosses, thin diamonds, diamonds, stars, and circles show the positions of NH$_3$ (6,3), (7,4), (8,5), (9,6), and (10,7) emissions. H$_2$O \citep{2004ApJS..155..577M}, class II CH$_3$OH \citep{1996MNRAS.283..606C,2019ApJS..244...35L}, H$_2$CO \citep{1994ApJ...434..237M,2007ApJ...654..971H,2019ApJS..244...35L}, and OH \citep{1990ApJ...351..538G} masers are presented as triangles, squares, pentagons, and hexagons, respectively. The color bar indicates the velocity range ($V_{\rm LSR}$) of maser spots. Red crosses mark the positions of the hot cores Sgr B2(N1), N2, and N3, taken from the 3 mm imaging line survey "Exploring Molecular Complexity with ALMA" (EMoCA, \citealt{2017A&A...604A..60B}). The systemic velocities of the hot cores N1, N2, and N3 are $V_{\rm LSR}$~=~64~km~s$^{-1}$, $V_{\rm LSR}$~=~74~km~s$^{-1}$ and $V_{\rm LSR}$~=~74~km~s$^{-1}$, respectively\citep{2017A&A...604A..60B}.}
    \label{sgrb2-k2_continuum_cm_mm}
\end{figure*}
\begin{figure*}[h]
\center
    \includegraphics[width=500pt]{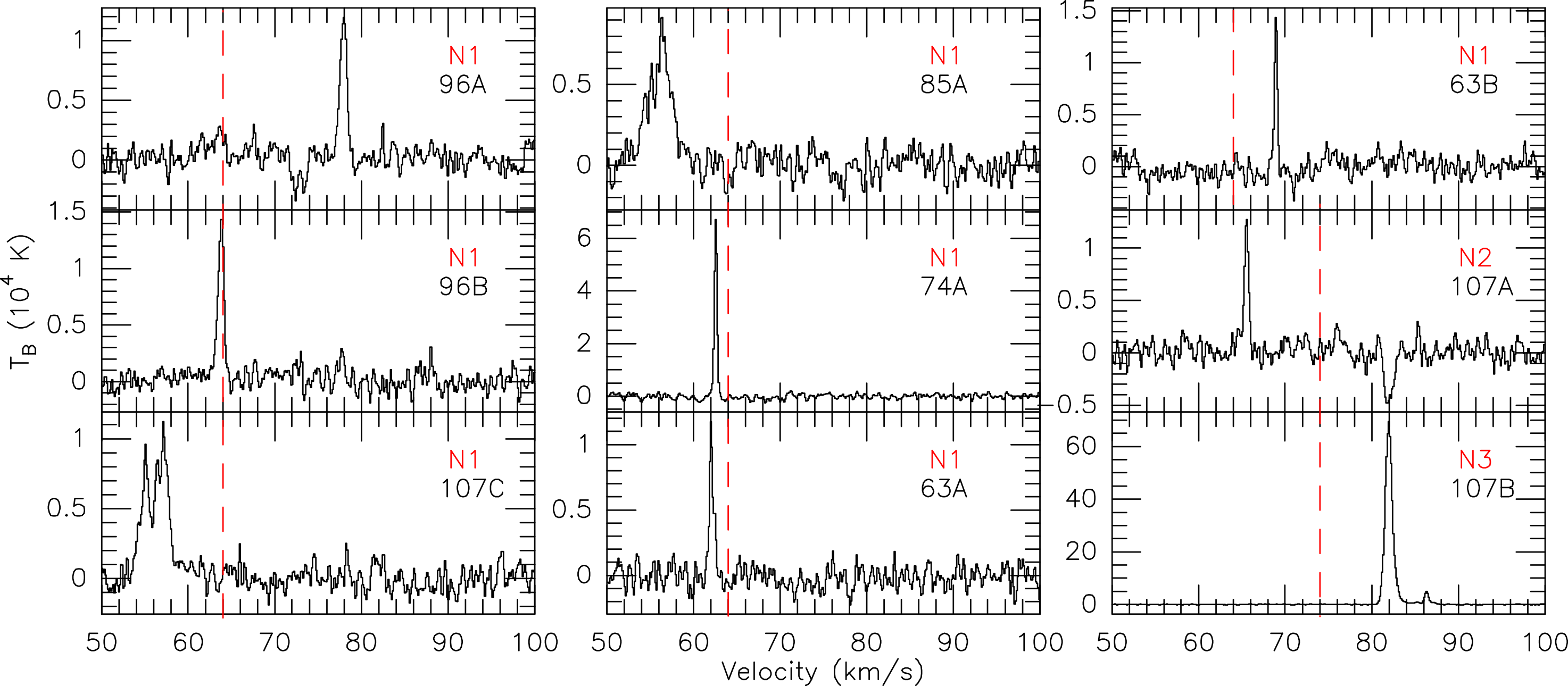}
  \caption{JVLA A-configuration spectra of NH$_3$ transition lines toward Sgr B2(N). The dashed red lines indicate the systemic velocities of the associated hot cores. $V_{\rm LSR}$~=~64~km~s$^{-1}$ for N1, and $V_{\rm LSR}$~=~74~km~s$^{-1}$ for N2 and N3 \citep{2017A&A...604A..60B}. Main beam brightness temperature scales are presented on the left hand side of the profiles.}
  \label{sgrb2-k2_spectra}
\end{figure*}

\textbf{Sgr B2(N)}: Among the 18 maser spots detected in Sgr B2, 50\% are located in Sgr B2(N). These are seen in the NH$_3$ (6,3) transition toward 63A and 63B, in the (7,4) transition toward 74A, in the (8,5) line toward 85A, in the (9,6) transition toward 96A and 96B, as well as in the (10,7) line toward 107A, 107B, and 107C. Four of these sources, 85A, 96A, 96B, and 107C, are close to the UC\h2 region K2 (see Fig.~\ref{sgrb2-k2_continuum_cm_mm}). The two maser spots, 85A and 107C, share the same position and have similar velocity distributions. Three maser spots, 63A, 63B, and 74A, surround the compact continuum source K3. The two maser spots, 63A and 74A, share, within the uncertainties, the same position and have similar velocity distributions. One maser spot, 107A, originates from a region offset by ($-0\farcs61\pm0\farcs02$, $+0\farcs47\pm0\farcs01$) from the continuum source K7. The second maser spot in the (10,7) line, 107B, has the highest brightness temperature. The lower limit is no less than 6$\times$10$^5$~K and originates from a region without any centimeter continuum source. Spectra from these nine maser spots are presented in Fig.~\ref{sgrb2-k2_spectra}. 

\textbf{Sgr B2(NS)}: Only one maser spot, 63C, in the NH$_3$ (6,3) transition was detected in this region. It is the strongest among all the detected (6,3) masers in Sgr B2 and is at a position with an offset of ($-0\farcs09\pm0\farcs02$, $+0\farcs01\pm0\farcs01$) from the \h2 region, Z10.24 (see Fig.~\ref{sgrb2-ns_continuum_cm} and Table~\ref{NH3-positions}). The spectrum is shown in Fig.~\ref{sgrb2-ns-spectra}. 

\textbf{Sgr B2(M)}: We detected five maser spots in this area (see Fig.~\ref{sgrb2-f10_continuum_cm_mm}). Three of them, 63D, 74B, and 96C, arise from similar positions close to the UC\h2 region F10.39. These three maser features are distributed in the same velocity range, from 70.0~km~s$^{-1}$ to 70.4~km~s$^{-1}$, while the (6,3) maser, 63D, extends spectroscopically down to the lower velocity of 68.0~km~s$^{-1}$. Spectra are presented in Fig.~\ref{sgrb2-m_spectra}. Another two maser spots, 85B and 107D, in the (8,5) and (10,7) transitions are located in regions close to the UC\h2 region F10.39.  

\textbf{Sgr B2(S)}: Three maser spots, 74C, 96D, and 107E, in the NH$_3$ (7,4), (9,6), and (10,7) transitions, are detected in this region. These are close to each other and are found slightly off the head of the cometary UC\h2 source (Fig.~\ref{sgrb2-s_continuum_cm_mm}). Spectra are shown in Fig.~\ref{sgrb2-s_spectra}.

\section{Discussion}
\label{disscussion}

As shown in Figs.~\ref{spectra-sgrb2n} and \ref{spectra-sgrb2m}, and claimed in Sect.~\ref{results}, our JVLA data of NH$_3$ (9,6) and (10,7) lines are not affected by missing flux. All of the detected maser spots are spatially unresolved, so the derived brightness temperatures are lower limits. Nevertheless, lower limits to the brightness temperature are $>$3000~K and reach 6$\times$10$^5$~K (107B). A comparison with the NH$_3$ (6,3), (7,4), (8,5), and (9,6) observations toward Sgr B2(N), using the TMRT 65-m telescope in March 2020 \citep{2020ApJ...898..157M}, reveals that almost all maser spots in our JVLA data are new detections at different velocities, with the exception of 74A. This strongly suggests substantial variations of the NH$_3$ (6,3), (8,5), and (9,6) masers since March 2020. 

Maser spots from other molecules, H$_2$O \citep{2004ApJS..155..577M}, class II CH$_3$OH \citep{1996MNRAS.283..606C,2016ApJ...833...18H,2019ApJS..244...35L}, H$_2$CO \citep{1994ApJ...434..237M,2007ApJ...654..971H,2019ApJS..244...35L}, and OH \citep{1990ApJ...351..538G}, are presented in Figs.~\ref{sgrb2-k2_continuum_cm_mm},~\ref{sgrb2-ns_continuum_cm},~\ref{sgrb2-f10_continuum_cm_mm}, and \ref{sgrb2-s_continuum_cm_mm}. The class I CH$_3$OH masers at 44~GHz toward the Sgr B2(M) and N regions from \citet{1997ApJ...474..346M} are outside the scales of Figs.~\ref{sgrb2-k2_continuum_cm_mm}, \ref{sgrb2-ns_continuum_cm}, and \ref{sgrb2-f10_continuum_cm_mm}, and are therefore not shown. The SiO maser \citep{1992PASJ...44..373M,2009ApJ...691..332Z} and the metastable NH$_3$ (2,2) masers in Sgr B2(M) \citep{2018ApJ...869L..14M} originate from a region around the radio object F3, which is more than two arcseconds north of source F10.39. These sources are, therefore, also not shown in Fig.~\ref{sgrb2-f10_continuum_cm_mm}. The ammonia (3,3) masers are located in an area more than 15 arcseconds south of Sgr B2(S) \citep{1999ApJ...519..667M} and outside the region shown in Fig.~\ref{sgrb2-s_continuum_cm_mm}. It is noteworthy that they are not spatially related to any (6,3) maser component, another transition within the same $K$ = 3 ladder. Our detected non-metastable NH$_3$ masers and previously detected metastable NH$_3$ masers arise from different regions. This indicates that these types of ammonia masers are excited in different ways.  

There are no apparent space and velocity correlations between our detected non-metastable NH$_3$ maser spots and other molecular masers. The locations of the hot cores Sgr B2(N1), N2 and N3 in Sgr B2(N), as well as Sgr B2(N5) in Sgr B2(NS), derived from the 3 mm imaging line survey "Exploring Molecular Complexity with ALMA" (EMoCA, \citealt{2017A&A...604A..60B}), are shown in Figs.~\ref{sgrb2-k2_continuum_cm_mm} and \ref{sgrb2-ns_continuum_cm}. A bipolar outflow in an east-west direction was found around the UC\h2 region K2, also known as Sgr B2(N1) \citep{2015ApJ...815..106H,2017A&A...604A..60B}. Bipolar outflows are also observed in a north-south direction and a northeast-southwest direction in the hot cores of Sgr B2 (N3) and N5, respectively \citep{2017A&A...604A..60B}. There is a class II CH$_3$OH maser spot and an H$_2$O maser spot close to the hot core Sgr B2(N3). Seven NH$_3$ maser spots, 63A, 63B, 74A, 85A, 96A, 96B, and 107C, are close to the hot core Sgr B2(N1). 107A and 107B originate from regions near the hot cores Sgr B2(N2) and N3, respectively. 63C arises from an area close to the hot core Sgr B2(N5). None of the NH$_3$ masers are spatially coincident with the hot cores in projection. The redshifts seen in 63B, 96A, and 107B, as well as blueshifts seen in 63A, 63C, 74A, 85A, 96B, 107A, and 107C with respect to the systemic velocities of the associated hot cores, may suggest that these maser spots are related to the outflows. The line profiles from these ten maser spots are shown in Figs.~\ref{sgrb2-k2_spectra} and \ref{sgrb2-ns-spectra}. \citet{2017A&A...604A..60B} did not find any sign of an outflow around Sgr B2(N2), while our maser spot 107A, with a blueshifted velocity of $V_{\rm LSR}$~=~65.5~km~s$^{-1}$ ($V_{\rm sys}$~=~74.0~km~s$^{-1}$), could indicate the presence of an outflow. 

All detected NH$_3$ maser spots in Sgr B2(M) show redshifted velocities with respect to the systemic velocity of Sgr B2(M), $V_{\rm LSR}$~=~62~km~s$^{-1}$ \citep{2013A&A...559A..47B}. This supports the suggestion that NH$_3$ masers in Sgr B2(M) are also related to outflows.

Toward Sgr B2(S), the ammonia maser spots 74C, 96D, and 107E also have redshifted velocities with respect to the systemic velocity of Sgr B2(S), $V_{\rm LSR}$~=~60~km~s$^{-1}$ \citep{2022arXiv220807796M}, which may indicate that this emission takes part in outflows. Several hot cores were identified by Jeff et al. (in prep.) in Sgr B2(S) in ALMA data (project code: 2017.1.00114.S, PI: A. Ginsburg). No hot cores were found to be close to the ammonia masers. This indicates that the presence of hot, dense gas alone is not sufficient to excite these masers. The ammonia masers detected in Sgr B2(S) are close to the head of the cometary UC\h2 region, similar to the NH$_3$ (9,6) maser M1 in G34.26$+$0.15 \citep{2022A&A...659A...5Y}. The maser spots in Sgr B2(S) show almost twice the angular distance compared to M1, with an offset of ($+0\farcs36\pm0\farcs13$, $-0\farcs07\pm0\farcs11$) in G34.26$+$0.15. In view of the different distances of G34.26+0.15 \citep[$D~\sim$~3.3~kpc][]{2022A&A...659A...5Y} and Sgr B2 (8.2~kpc see Sect.\ref{introduction}), the linear distance is even five times larger in Sgr B2(S), that is 0.03 pc. The velocity difference between the masers and the cometary UC\h2 regions is ten times higher in Sgr B2(S) than in G34.26+0.15 ($\Delta V_{\rm SgrB2(S)}\geq$17.4~km~s$^{-1}$ and $\Delta V_{\rm G34.26}\sim$-1.3~km~s$^{-1}$). That indicates that the cometary UC\h2 region in Sgr B2(S) is more active than the one in G34.26$+$0.15.

Overall, the detected non-metastable ammonia masers in Sgr B2 are consistent with the discussion on pumping scenarios in \citet{2022A&A...659A...5Y}. Therefore, we speculate that the detected NH$_3$ masers in the non-metastable (6,3), (7,4), (8,5), (9,6), and (10,7) transitions in Sgr B2(N), Sgr B2(NS), Sgr B2(M), and Sgr B2(S) appear to be associated with shocks caused either by outflows or UCH{\scriptsize II} expansion.

\section{Summary}
\label{summary}

We report the discovery of NH$_3$ non-metastable (6,3), (7,4), (8,5), (9,6), and (10,7) masers in Sgr B2(M) and Sgr B2(N), an NH$_3$ (6,3) maser in Sgr B2(NS), as well as NH$_3$ (7,4), (9,6), and (10,7) masers in Sgr B2(S). High angular resolution data from the JVLA A-configuration reveal 18 maser spots. Nine maser spots arise from Sgr B2(N), one from Sgr B2(NS), five from Sgr B2(M), and three originate in Sgr B2(S). All of these increase the number of (6,3), (7,4), (8,5), (9,6), and (10,7) maser detections in our Galaxy from three to six, two to four, two to three, seven to nine, and one to four. Compared to the Effelsberg 100-m telescope data, the JVLA data indicate no missing flux. The detected maser spots are not resolved by our JVLA observations. Lower limits to the brightness temperature are $>$3000~K and reach up to 6$\times$10$^5$~K, manifesting their maser nature. Long-term Effelsberg monitoring (19 months) indicates that the intensities of the (9,6) masers in Sgr B2(M), as well as the (9,6) and (10,7) masers in Sgr B2(N), show noticeable variations. However, the (10,7) maser in Sgr B2(M) is stable. While the NH$_3$ masers all arise near hot cores, there are many hot cores that do not exhibit NH$_3$ maser emission. All of these non-metastable ammonia maser lines show redshifted or blueshifted features that may be related to outflows or UCH{\scriptsize II} expansion.

\begin{acknowledgements}
The authors thank the anonymous referee for the useful comments that improve the manuscript. We thank Chris De Pree for providing the 7 mm continuum images of Sgr B2(M) and Sgr B2(N). Y.T.Y. is a member of the International Max Planck Research School (IMPRS) for Astronomy and Astrophysics at the Universities of Bonn and Cologne. Y.T.Y. thanks the China Scholarship Council (CSC) and the Max-Planck-Institut f\"{u}r Radioastronomie (MPIfR) for the financial support. Y.T.Y. also thanks his fiancee, Siqi Guo, for her support during this pandemic period. We would like to thank the staff at the Effelsberg telescope for their help provided during the observations. The National Radio Astronomy Observatory is a facility of the National Science Foundation operated
under cooperative agreement by Associated Universities, Inc. We thank the staff of the JVLA, especially Tony Perreault and Drew Medlin, for their assistance with the observations and data reduction.
\end{acknowledgements}

\bibliographystyle{aa}
\bibliography{ammoniamaser}

\begin{appendix}
\label{appendix}
\onecolumn

\section{Figures}
\label{appendix-figures}

\begin{figure}[h]
\center
    \includegraphics[width=280pt]{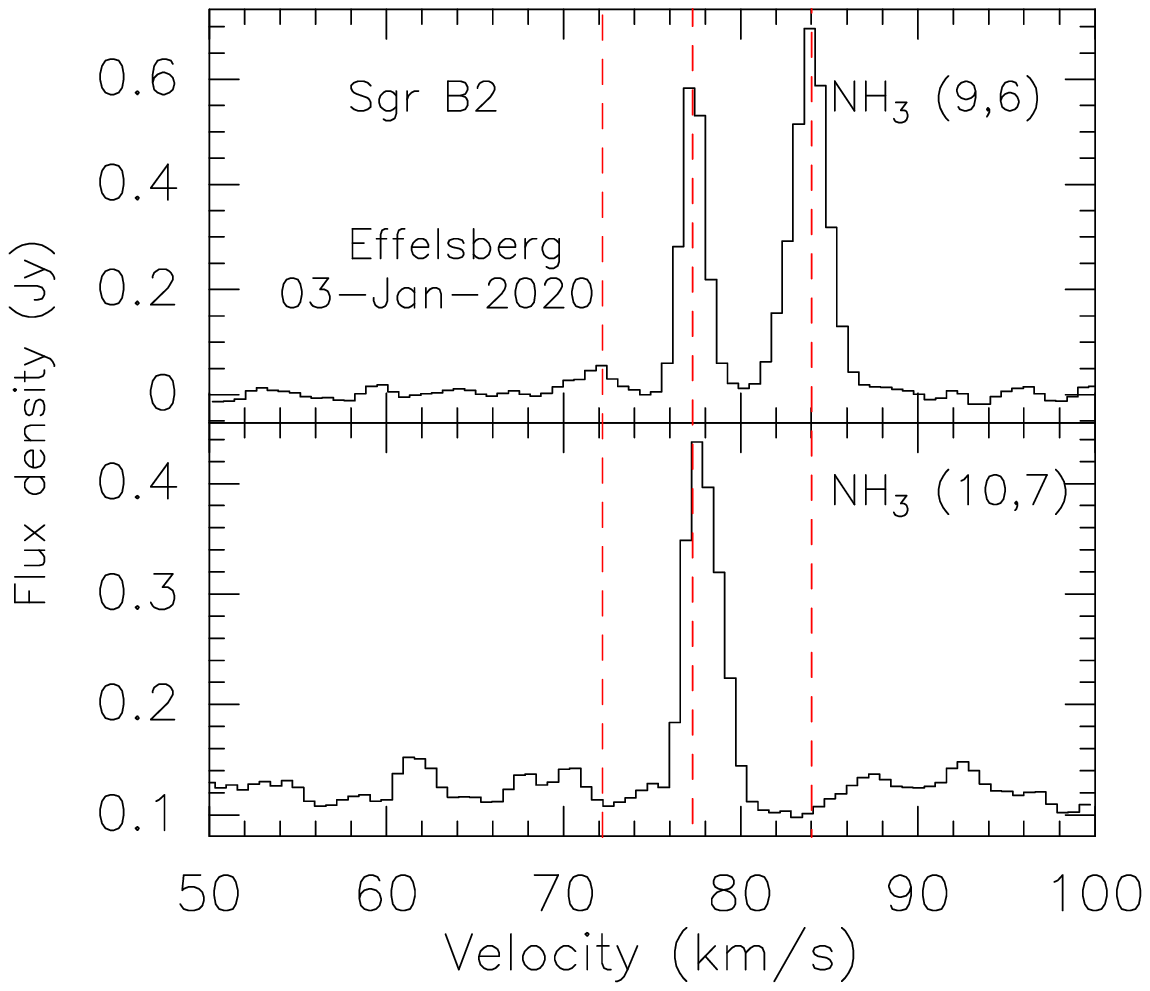}
  \caption{\label{spectra-2020} NH$_3$ (9,6) and (10,7) maser lines from the 100-m telescope at Effelsberg toward a region south of Sgr B2(M), at $\alpha_{\rm J2000}$~=~17$^{\rm h}$47$^{\rm m}$20$\fs$8, $\delta_{\rm J2000}$~=~$-$28$\degr$23$\arcmin$32$\farcs$1. The three red dashed lines indicate the three different velocity components at $V_{\rm LSR}$~=~$\sim$72 km s$^{-1}$, $\sim$77 km s$^{-1}$, and $\sim$84 km~s$^{-1}$.}
\end{figure}

\begin{figure}[h]
\center
    \includegraphics[width=500pt]{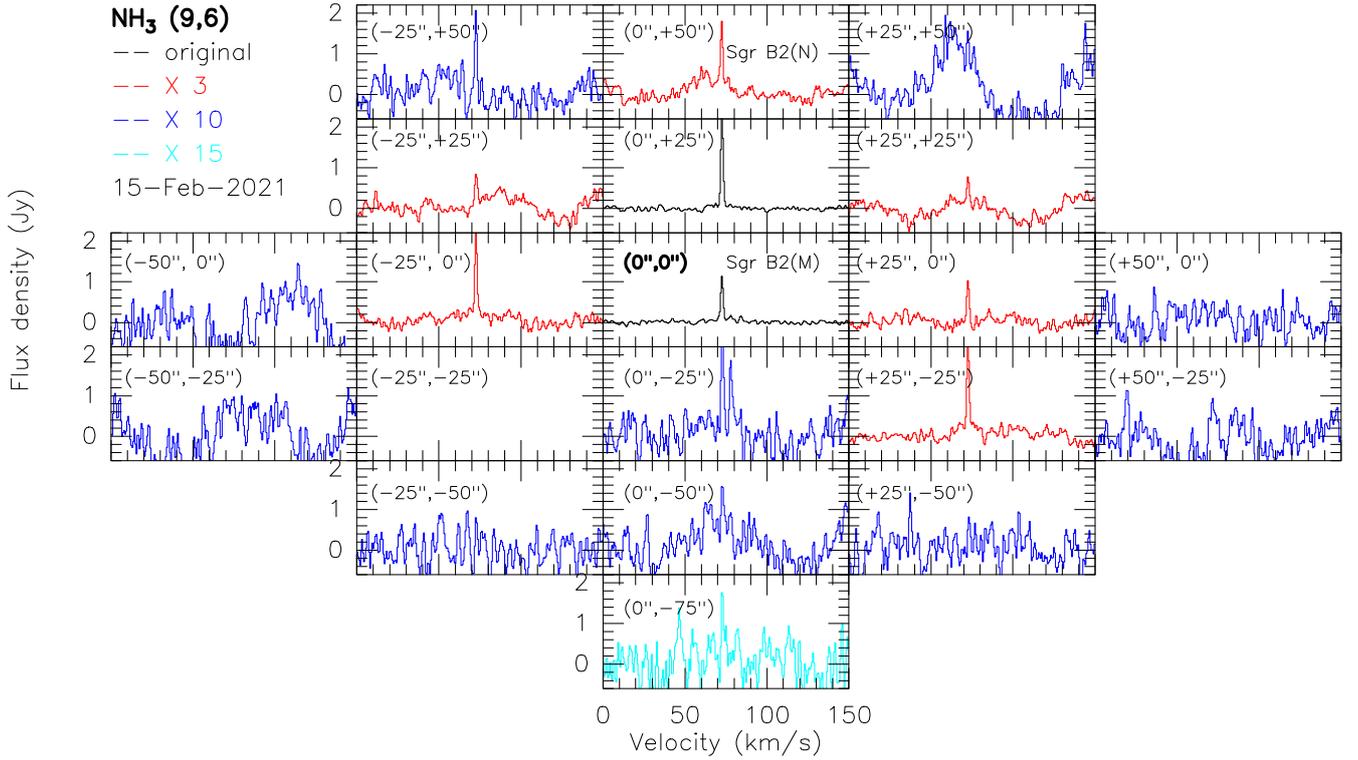}
  \caption{\label{96spectra-2021} NH$_3$ (9,6) line profiles observed with the Effelsberg 100-m telescope over the region of Sgr B2. The black spectra show the original flux density scales. The red, blue, and cyan spectra are presented after multiplying the flux densities by factors of three, ten, and fifteen, respectively. }
\end{figure}

\begin{figure}[h]
\center
    \includegraphics[width=500pt]{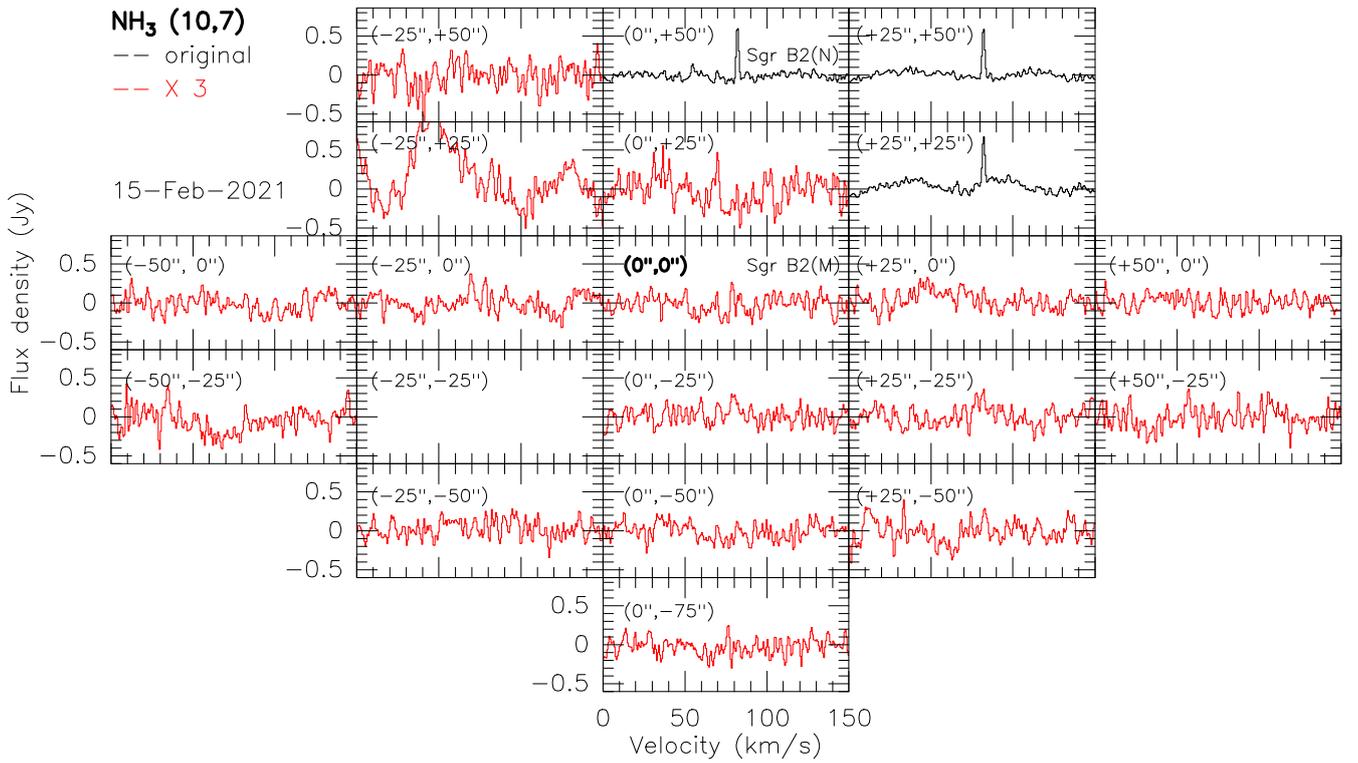}
  \caption{\label{107spectra-2021} NH$_3$ (10,7) spectra observed with the Effelsberg 100-m telescope toward Sgr B2. The black spectra show the original flux density scales, while the red spectra are presented after multiplying the flux densities by a factor of three. }
\end{figure}

\begin{figure}[h]
\center
    \includegraphics[height=400pt]{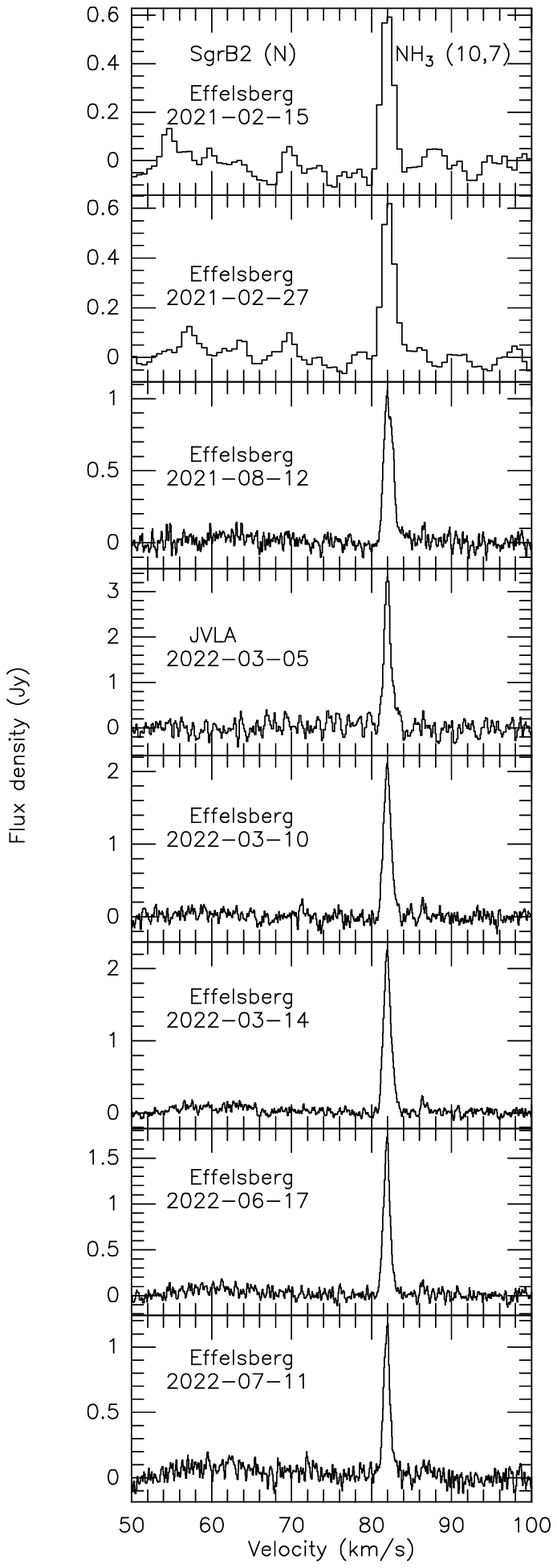}
    \includegraphics[height=400pt]{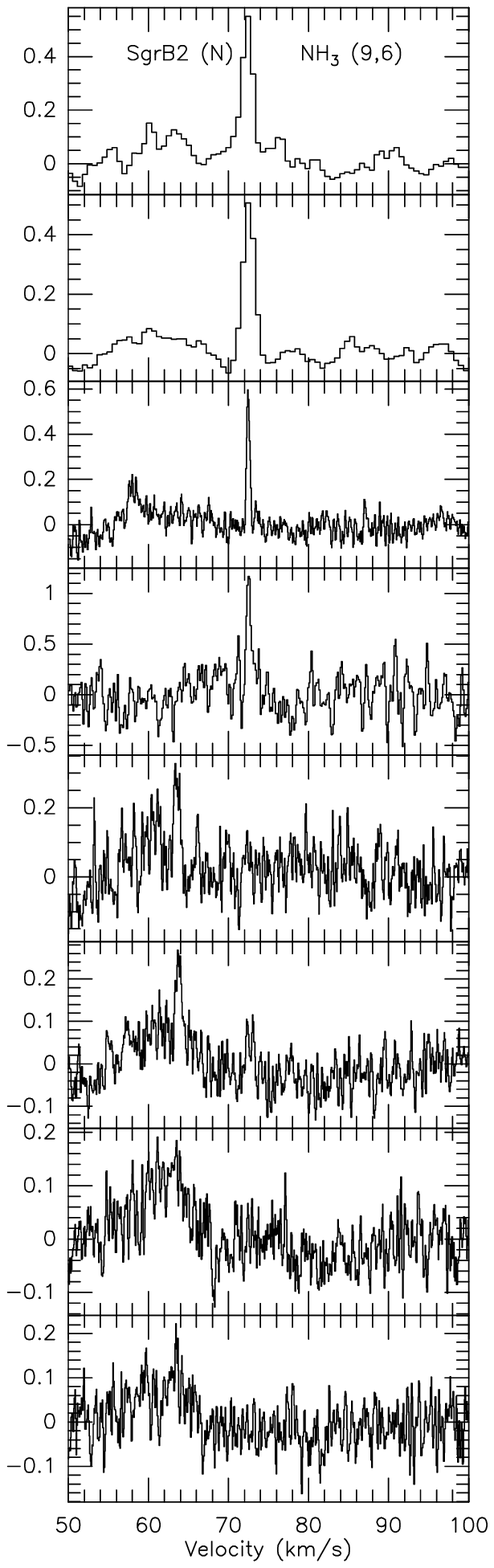}
  \caption{\label{spectra-sgrb2n} Effelsberg 100-meter telescope and JVLA A-configuration spectra from NH$_3$ (9,6) and (10,7) transition lines at eight epochs toward Sgr B2(N), after subtracting a first-order polynomial baseline. The JVLA spectra are extracted over a region of radius 35$\arcsec$ centered at Sgr B2(N).}
\end{figure}

\begin{figure}[h]
\center
    \includegraphics[height=600pt]{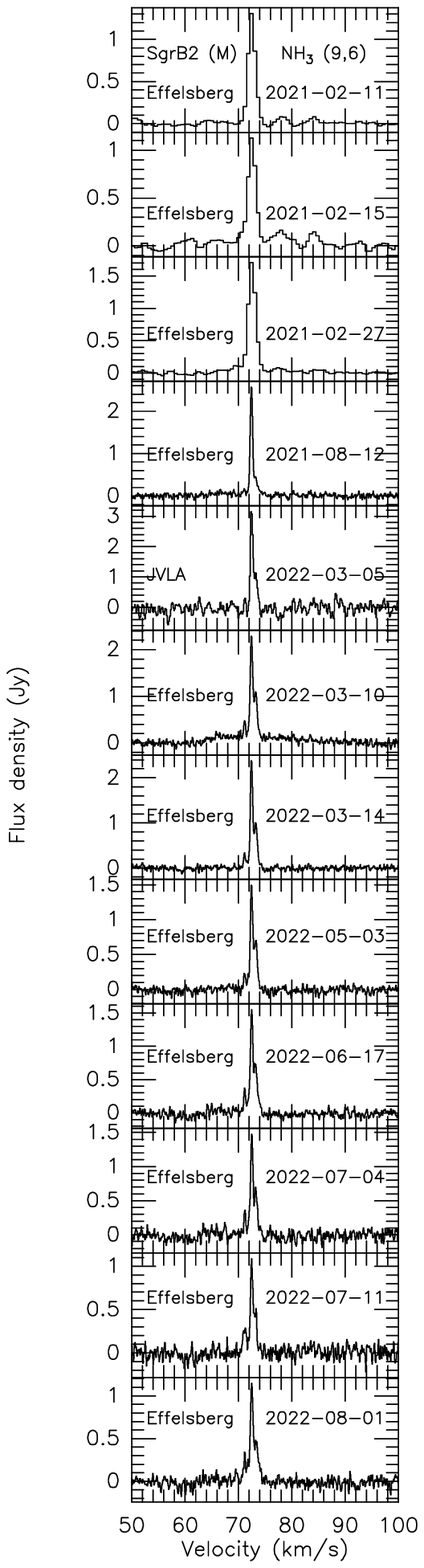}
    \includegraphics[height=600pt]{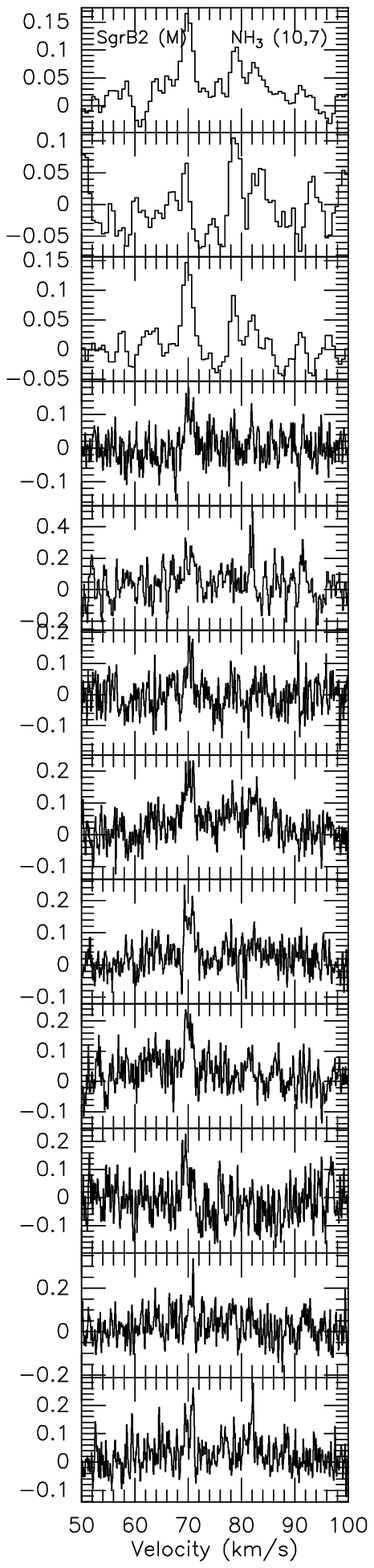}
  \caption{\label{spectra-sgrb2m} Effelsberg 100-meter telescope and JVLA A-configuration spectra from NH$_3$ (9,6) and (10,7) transition lines at 12 epochs toward Sgr B2(M), after subtracting a first-order polynomial baseline. The JVLA spectra are extracted from the Effelsberg beam (FWHM, 49$\arcsec$) sized region.}
\end{figure}

\twocolumn
\begin{figure}[h]
\center
    \includegraphics[width=270pt]{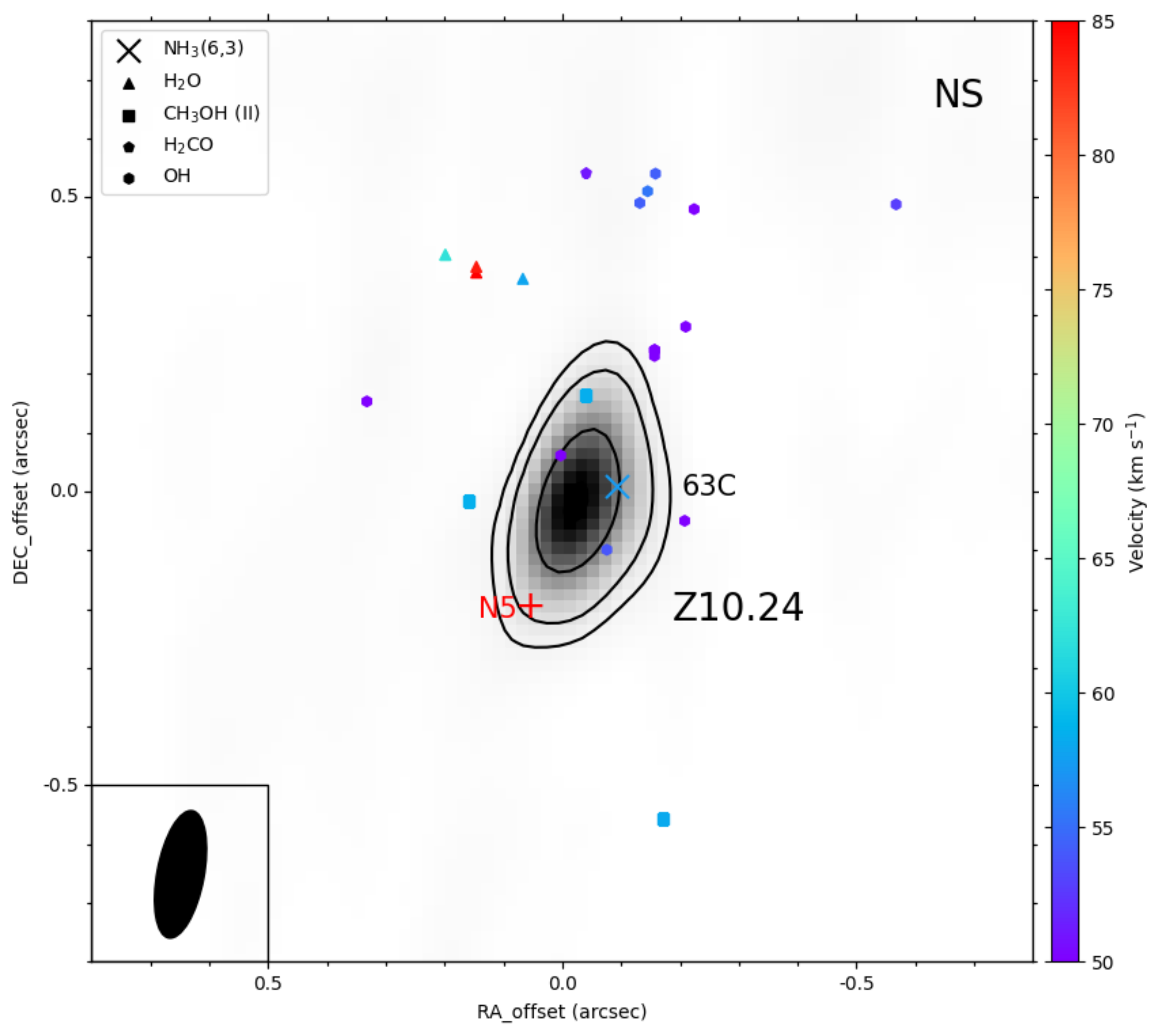}
    \caption{JVLA 1.6\,cm continuum map of Sgr B2(NS) presented as gray shaded area and black contours with levels of 5, 10, 30, and 50~$\times$~0.2~mJy~beam$^{-1}$. The reference position is $\alpha_{\rm J2000}$~=~17$^{\rm h}$47$^{\rm m}$20$\fs$043, and $\delta_{\rm J2000}$~=~$-$28$\degr$22$\arcmin$41$\farcs$143, the peak position of continuum source Z10.24. The yellow cross shows the position of NH$_3$ (6,3) emission. H$_2$O \citep{2004ApJS..155..577M}, class II CH$_3$OH \citep{1996MNRAS.283..606C,2016ApJ...833...18H,2019ApJS..244...35L}, H$_2$CO \citep{2007ApJ...654..971H}, and OH \citep{1990ApJ...351..538G} masers are presented as triangles, squares, pentagons, and hexagons, respectively. The color bar indicates the velocity range ($V_{\rm LSR}$) of the maser spots. The red cross marks the position of the hot core, Sgr B2(N5), taken from the 3 mm imaging line survey EMoCA (\citealt{2017A&A...604A..60B}). The systemic velocity of the hot core, N5, is $V_{\rm LSR}$~=~60~km~s$^{-1}$ \citep{2017A&A...604A..60B}.}
    \label{sgrb2-ns_continuum_cm}
\end{figure}

\begin{figure}[h]
\center
    \includegraphics[width=270pt]{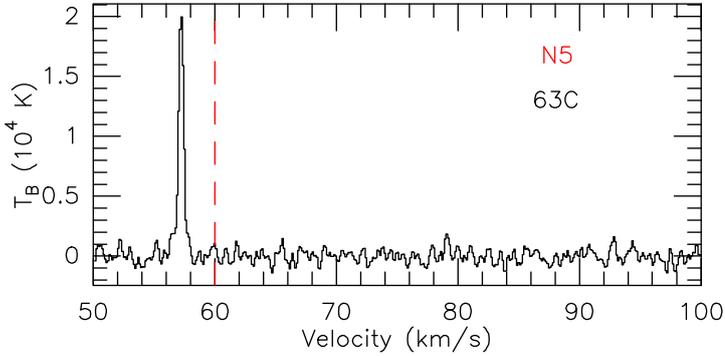}
  \caption{JVLA A-configuration spectrum of the NH$_3$ (6,3) transition line toward Sgr B2(NS). The systemic velocity of the associated hot core, $V_{\rm LSR}$~=~60~km~s$^{-1}$ in N5 \citep{2017A&A...604A..60B}, is indicated by the dashed red line.}
  \label{sgrb2-ns-spectra}
\end{figure}

\begin{figure}[h]
\center
    \includegraphics[width=270pt]{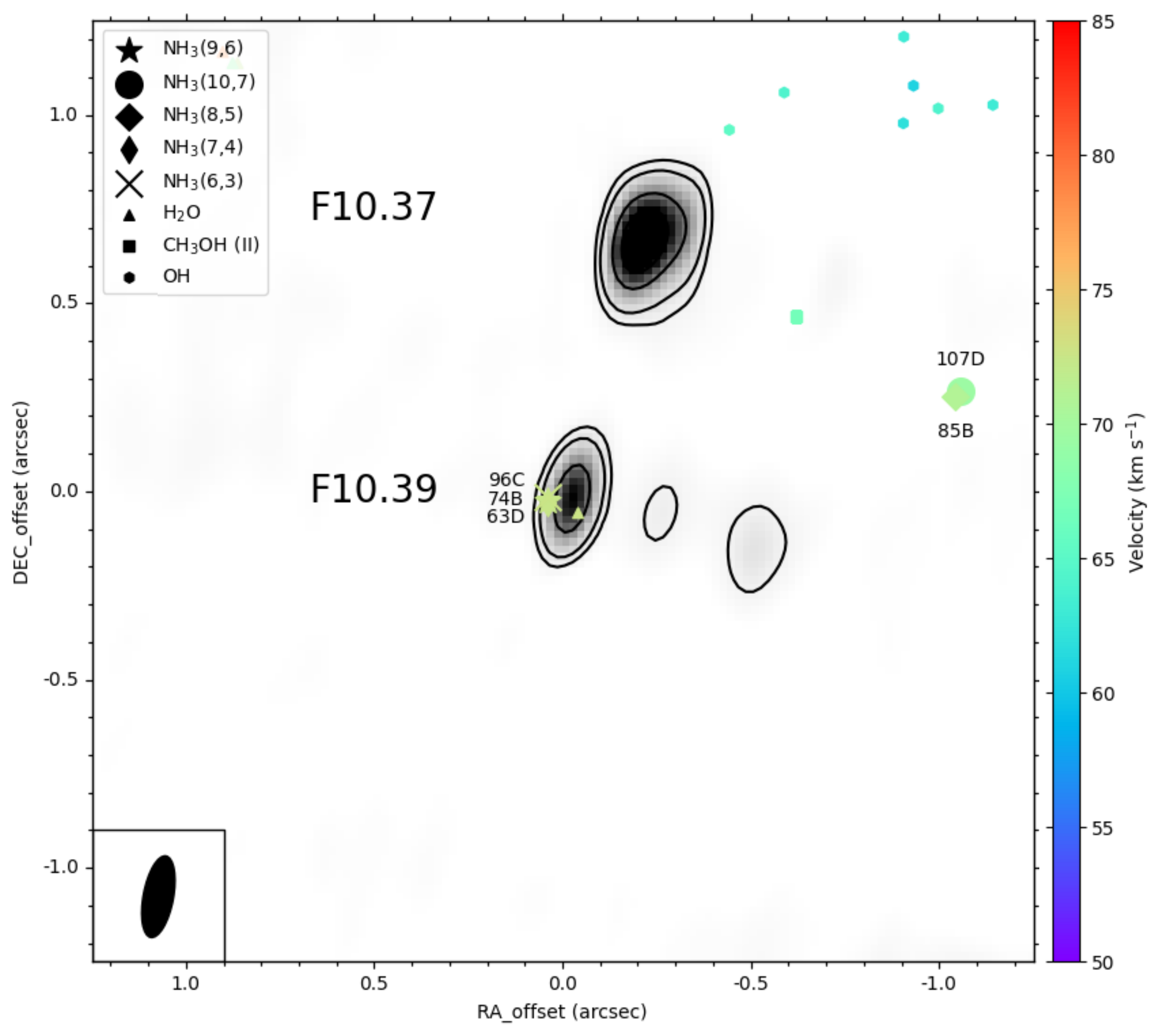}
    \caption{JVLA 1.6\,cm continuum map of Sgr B2(M) presented as gray shaded areas and black contours with levels of 5, 10, 30, and 50~$\times$~0.2~mJy~beam$^{-1}$. The reference position is $\alpha_{\rm J2000}$~=~17$^{\rm h}$47$^{\rm m}$20$\fs$197, and $\delta_{\rm J2000}$~=~$-$28$\degr$23$\arcmin$06$\farcs$484, the peak position of continuum source F10.39. The cross, thin diamond, diamond, circle and star show the positions of NH$_3$ (6,3), (7,4), (8,5), (9,6), and (10,7) emissions. H$_2$O \citep{2004ApJS..155..577M}, class II CH$_3$OH \citep{2019ApJS..244...35L}, and OH \citep{1990ApJ...351..538G} masers are presented as triangles, squares, and hexagons, respectively. The color bar indicates the velocity range ($V_{\rm LSR}$) of the maser spots. The systemic velocity of Sgr B2(M) is $V_{\rm LSR}$~=~62~km~s$^{-1}$ \citep{2013A&A...559A..47B}.}
    \label{sgrb2-f10_continuum_cm_mm}
\end{figure}

\begin{figure}[h]
\center
    \includegraphics[width=230pt]{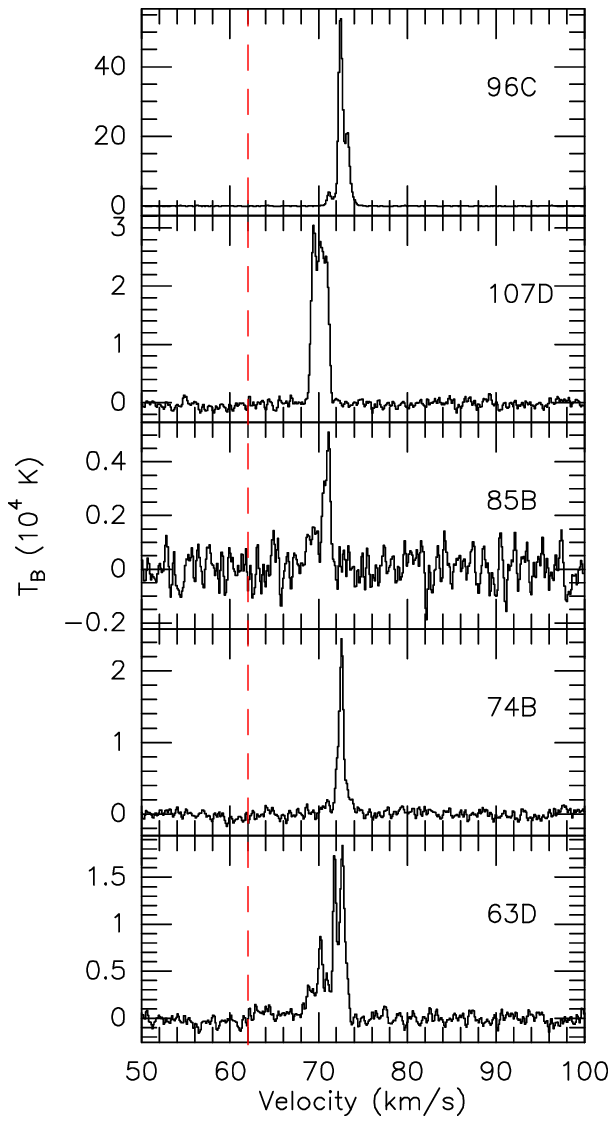}
  \caption{JVLA A-configuration spectra of NH$_3$ transition lines toward Sgr B2(M). The systemic velocity of Sgr B2(M), $V_{\rm LSR}$~=~62~km~s$^{-1}$ \citep{2013A&A...559A..47B}, is indicated by the dashed red line.}
  \label{sgrb2-m_spectra}
\end{figure}

\begin{figure}[h]
\center
    \includegraphics[width=270pt]{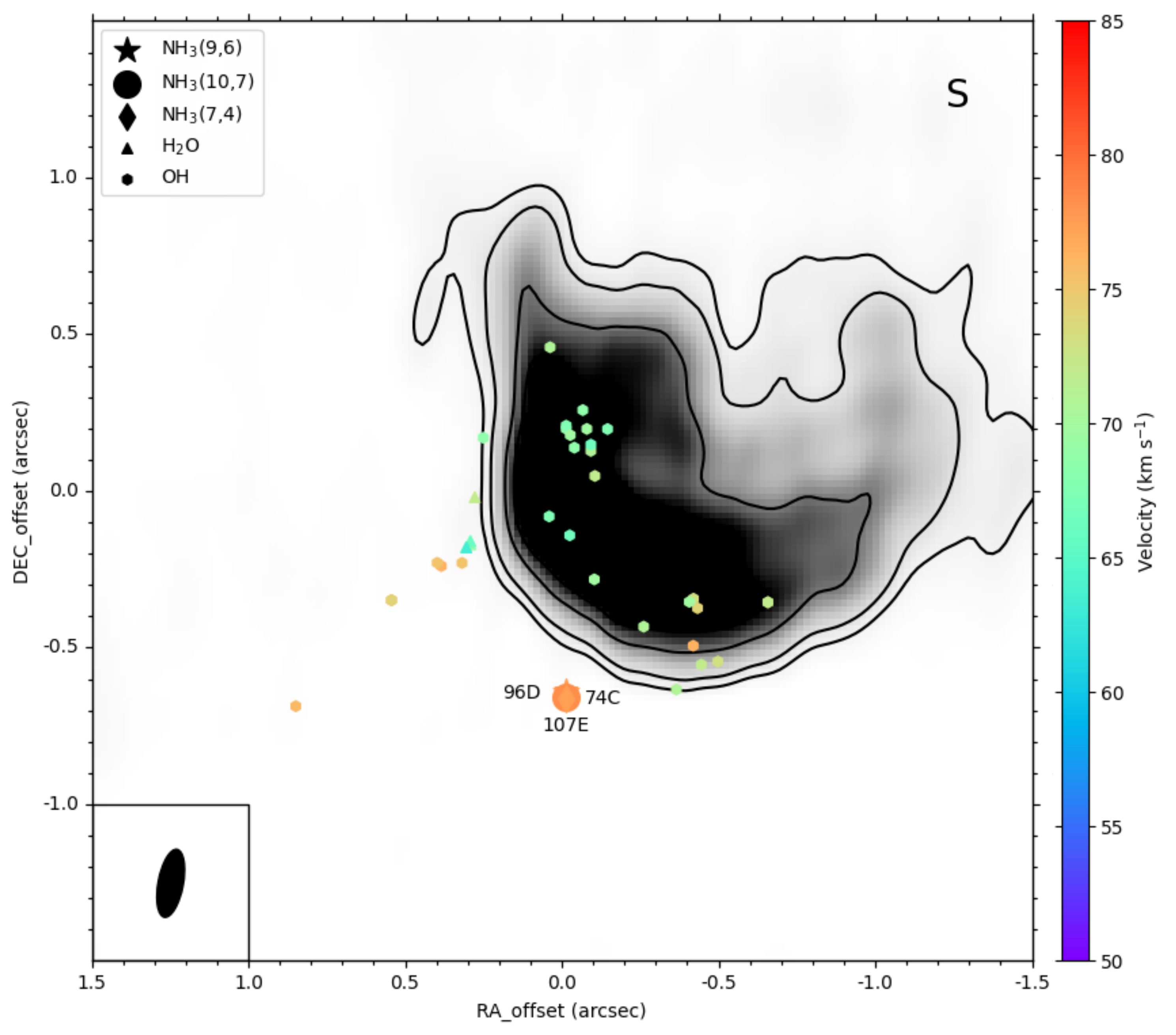}
    \caption{JVLA 1.6\,cm continuum map of Sgr B2(S) presented as gray shaded areas and black contours with levels of 5, 10, 30, and 50~$\times$~0.2~mJy~beam$^{-1}$. The reference position is $\alpha_{\rm J2000}$~=~17$^{\rm h}$47$^{\rm m}$20$\fs$472, and $\delta_{\rm J2000}$~=~$-$28$\degr$23$\arcmin$45$\farcs$120, the peak position of the continuum source. The thin diamond, star, and circle show the positions of NH$_3$ (7,4), (9,6), and (10,7) emissions. H$_2$O \citep{2004ApJS..155..577M} and OH \citep{1990ApJ...351..538G} masers are presented as triangles and hexagons, respectively. The color bar indicates the velocity range ($V_{\rm LSR}$) of the maser spots. The systemic velocity of Sgr B2(S) is $V_{\rm LSR}$~=~60~km~s$^{-1}$ \citep{2022arXiv220807796M}.}
    \label{sgrb2-s_continuum_cm_mm}
\end{figure}

\begin{figure}[h]
\center
    \includegraphics[width=230pt]{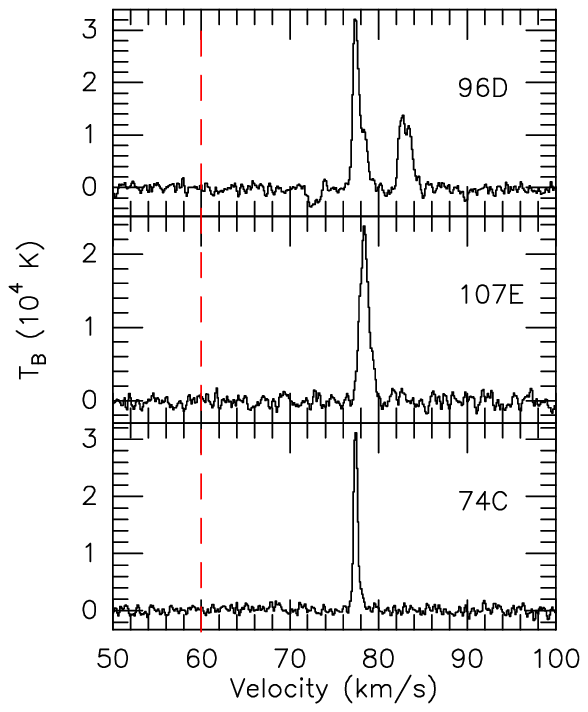}
  \caption{JVLA A-configuration spectra of NH$_3$ transition lines toward Sgr B2(S). The systemic velocity of Sgr B2(S), $V_{\rm LSR}$~=~60~km~s$^{-1}$ \citep{2022arXiv220807796M}, is indicated by the dashed red line.}
  \label{sgrb2-s_spectra}
\end{figure}

\onecolumn
\newpage
\section{Tables}
\label{appendix-tables}
\begin{table*}[h]
\caption{Summary of NH$_3$ (9,6) maser observations toward Sgr B2(N)}
\centering
\begin{tabular}{lccccccccc}
\hline\hline
Source & Telescope & Beam & Epoch &  Channel  & $S_\nu$  & rms & $\int S_\nu dv$ & $V_{\rm LSR}$ & $\Delta V_{1/2}$  \\
  &   & size & & spacing  &    &  & & &   \\
  &   &  & & (km s$^{-1}$) & (Jy) &  (Jy)  & (Jy km s$^{-1}$) & \multicolumn{2}{c}{ (km s$^{-1}$)}    \\
\hline
\label{spectra96_fitting-sgrb2n}
Sgr B2(N)  & Effelsberg & 49$\arcsec$ & 2021, Feb. 15  & 0.62 & 0.53 & 0.034  & 1.03 $\pm$ 0.09 & 72.34 $\pm$ 0.07 & 1.83 $\pm$ 0.21 \\
           & Effelsberg & 49$\arcsec$ & 2021, Feb. 27  & 0.62 & 0.52 & 0.028  & 0.92 $\pm$ 0.06 & 72.54 $\pm$ 0.05 & 1.65 $\pm$ 0.12 \\
           & Effelsberg & 49$\arcsec$ & 2021, Aug. 12  & 0.07 & 0.09 & 0.043  & 1.16 $\pm$ 0.09 & 63.02 $\pm$ 0.50 & 12.14 $\pm$ 0.88 \\
           &            &             &                &      & 0.14 &        & 0.34 $\pm$ 0.06 & 58.09 $\pm$ 0.09 & 2.17 $\pm$ 0.39 \\
           &            &             &                &      & 0.61 &        & 0.30 $\pm$ 0.01 & 72.48 $\pm$ 0.01 & 0.47 $\pm$ 0.02 \\
  & JVLA\tablefootmark{a}& $0\farcs215~\times~0\farcs097$ & 2022, Mar. 05  & 0.13 & 1.16 & 0.189  & 0.89 $\pm$ 0.09 & 72.55 $\pm$ 0.04 & 0.72 $\pm$ 0.10 \\
           & Effelsberg & 49$\arcsec$ & 2022, Mar. 10  & 0.07 & 0.30 & 0.067 & 0.32 $\pm$ 0.03  & 63.52 $\pm$ 0.05 & 1.01 $\pm$ 0.12 \\
           & Effelsberg & 49$\arcsec$ & 2022, Mar. 14  & 0.07 & 0.12 & 0.041 & 1.37 $\pm$ 0.06  & 60.99 $\pm$ 0.22 & 10.29 $\pm$ 0.47 \\
           &            &             &                &      & 0.20 &       & 0.16 $\pm$ 0.02  & 63.81 $\pm$ 0.04 & 0.79 $\pm$ 0.09 \\
           &            &             &                &      & 0.10 &       & 0.12 $\pm$ 0.02  & 72.66 $\pm$ 0.09 & 1.12 $\pm$ 0.15 \\
           & Effelsberg & 49$\arcsec$ & 2022, Jun. 17  & 0.07 & $\cdots$ & 0.041 & $\cdots$ & $\cdots$ & $\cdots$ \\
           & Effelsberg & 49$\arcsec$ & 2022, Jul. 11  & 0.07 & $\cdots$ & 0.047 & $\cdots$ & $\cdots$ & $\cdots$ \\
\hline
\end{tabular}
\tablefoot{
The spectral parameters are obtained from Gaussian-fitting. \tablefoottext{a}{The JVLA spectrum is extracted from a region of radius 35$\arcsec$.} }
\end{table*}

\begin{table*}[h]
\caption{Summary of NH$_3$ (10,7) maser observations toward Sgr B2(N)}
\centering
\begin{tabular}{lccccccccc}
\hline\hline
Source & Telescope & Beam & Epoch &  Channel  & $S_\nu$  & rms & $\int S_\nu dv$ & $V_{\rm LSR}$ & $\Delta V_{1/2}$  \\
  &   & size & & spacing  &    &  & & &   \\
  &   &  & & (km s$^{-1}$) & (Jy) &  (Jy)  & (Jy km s$^{-1}$) & \multicolumn{2}{c}{ (km s$^{-1}$)}    \\
\hline
\label{spectra107_fitting-sgrb2n}
Sgr B2(N)  & Effelsberg & 49$\arcsec$ & 2021, Feb. 15  & 0.62 & 0.66 & 0.048  & 1.12 $\pm$ 0.07 & 81.99 $\pm$ 0.05 & 1.60 $\pm$ 0.11 \\
           & Effelsberg & 49$\arcsec$ & 2021, Feb. 27  & 0.62 & 0.64 & 0.038  & 1.19 $\pm$ 0.06 & 82.10 $\pm$ 0.04 & 1.76 $\pm$ 0.10 \\
           & Effelsberg & 49$\arcsec$ & 2021, Aug. 12  & 0.07 & 1.02 & 0.045  & 1.44 $\pm$ 0.02 & 82.10 $\pm$ 0.01 & 1.33 $\pm$ 0.02 \\
  & JVLA\tablefootmark{a}& $0\farcs220~\times~0\farcs094$ & 2022, Mar. 05  & 0.13 & 3.21 & 0.141  & 3.49 $\pm$ 0.09 & 81.99 $\pm$ 0.01 & 1.02 $\pm$ 0.03 \\
           & Effelsberg & 49$\arcsec$ & 2022, Mar. 10  & 0.07 & 2.02 & 0.070 & 2.35 $\pm$ 0.03  & 81.97 $\pm$ 0.01 & 1.09 $\pm$ 0.02 \\
           & Effelsberg & 49$\arcsec$ & 2022, Mar. 14  & 0.07 & 2.12 & 0.051 & 2.48 $\pm$ 0.02  & 81.96 $\pm$ 0.01 & 1.10 $\pm$ 0.01 \\
           & Effelsberg & 49$\arcsec$ & 2022, Jun. 17  & 0.07 & 1.61 & 0.051 & 1.66 $\pm$ 0.02  & 81.90 $\pm$ 0.01 & 0.97 $\pm$ 0.02 \\
           & Effelsberg & 49$\arcsec$ & 2022, Jul. 11  & 0.07 & 1.11 & 0.055 & 1.13 $\pm$ 0.03  & 81.91 $\pm$ 0.01 & 0.96 $\pm$ 0.03 \\
\hline
\end{tabular}
\tablefoot{
The spectral parameters are obtained from Gaussian-fitting. \tablefoottext{a}{The JVLA spectrum is extracted from a region of radius 35$\arcsec$.}}
\end{table*}

\begin{table*}[h]
\caption{Summary of NH$_3$ (9,6) maser observations toward Sgr B2(M)}
\centering
\begin{tabular}{lccccccccc}
\hline\hline
Source & Telescope & Beam & Epoch &  Channel  & $S_\nu$  & rms & $\int S_\nu dv$ & $V_{\rm LSR}$ & $\Delta V_{1/2}$  \\
  &   & size & & spacing  &    &  & & &   \\
  &   &  & & (km s$^{-1}$) & (Jy) &  (Jy)  & (Jy km s$^{-1}$) & \multicolumn{2}{c}{ (km s$^{-1}$)}    \\
\hline
\label{spectra96_fitting-sgrb2m}
Sgr B2(M)  & Effelsberg & 49$\arcsec$ & 2021, Feb. 11  & 0.62 & 1.29 & 0.021  & 2.22 $\pm$ 0.03 & 72.49 $\pm$ 0.01 & 1.61 $\pm$ 0.03  \\
           &            &             &                &      & 0.09 &        & 0.17 $\pm$ 0.03 & 78.23 $\pm$ 0.17 & 1.76 $\pm$ 0.32 \\
           &            &             &                &      & 0.08 &        & 0.15 $\pm$ 0.04 & 84.05 $\pm$ 0.20 & 1.83 $\pm$ 0.51 \\
           & Effelsberg & 49$\arcsec$ & 2021, Feb. 15  & 0.62 & 1.10 & 0.035  & 2.13 $\pm$ 0.06 & 72.52 $\pm$ 0.02 & 1.82 $\pm$ 0.06 \\
           &            &             &                &      & 0.15 &        & 0.59 $\pm$ 0.08 & 77.83 $\pm$ 0.25 & 3.78 $\pm$ 0.59 \\
           &            &             &                &      & 0.15 &        & 0.30 $\pm$ 0.06 & 84.18 $\pm$ 0.19 & 1.84 $\pm$ 0.40 \\
           & Effelsberg & 49$\arcsec$ & 2021, Feb. 27  & 0.62 & 1.69 & 0.018  & 3.06 $\pm$ 0.05 & 72.53 $\pm$ 0.01 & 1.70 $\pm$ 0.03 \\
           & Effelsberg & 49$\arcsec$ & 2021, Aug. 12  & 0.07 & 2.30 & 0.040  & 1.17 $\pm$ 0.03 & 72.47 $\pm$ 0.01 & 0.48 $\pm$ 0.01 \\
           &            &             &                &      & 0.39 &        & 0.80 $\pm$ 0.04 & 72.92 $\pm$ 0.04 & 1.92 $\pm$ 0.11 \\
  & JVLA\tablefootmark{a}& $0\farcs215~\times~0\farcs097$ & 2022, Mar. 05  & 0.13 & 3.20 & 0.154  & 1.71 $\pm$ 0.15 & 72.45 $\pm$ 0.02 & 0.50 $\pm$ 0.03 \\
           &            &             &                &                          & 1.11 &      & 0.99 $\pm$ 0.16 & 73.19 $\pm$ 0.07 & 0.84 $\pm$ 0.13 \\
           & Effelsberg & 49$\arcsec$ & 2022, Mar. 10  & 0.07 & 0.39 & 0.048 & 0.33 $\pm$ 0.02  & 71.22 $\pm$ 0.07 & 0.80 $\pm$ 0.07 \\
           &            &             &                &      & 2.16 &       & 1.31 $\pm$ 0.02  & 72.45 $\pm$ 0.07 & 0.57 $\pm$ 0.07 \\
           &            &             &                &      & 1.00 &       & 0.91 $\pm$ 0.02  & 73.27 $\pm$ 0.07 & 0.85 $\pm$ 0.07 \\
           & Effelsberg & 49$\arcsec$ & 2022, Mar. 14  & 0.07 & 0.26 & 0.040 & 0.21 $\pm$ 0.02  & 71.26 $\pm$ 0.03 & 0.76 $\pm$ 0.07 \\
           &            &             &                &      & 2.38 &       & 1.25 $\pm$ 0.02  & 72.46 $\pm$ 0.01 & 0.49 $\pm$ 0.01 \\
           &            &             &                &      & 0.97 &       & 0.77 $\pm$ 0.02  & 73.26 $\pm$ 0.01 & 0.74 $\pm$ 0.03 \\
           & Effelsberg & 49$\arcsec$ & 2022, May. 03  & 0.07 & 0.18 & 0.037 & 0.20 $\pm$ 0.02  & 71.36 $\pm$ 0.06 & 1.01 $\pm$ 0.12 \\
           &            &             &                &      & 1.45 &       & 0.78 $\pm$ 0.03  & 72.47 $\pm$ 0.01 & 0.51 $\pm$ 0.02 \\
           &            &             &                &      & 0.70 &       & 0.56 $\pm$ 0.03  & 73.25 $\pm$ 0.02 & 0.75 $\pm$ 0.04 \\
           & Effelsberg & 49$\arcsec$ & 2022, Jun. 17  & 0.07 & 0.28 & 0.044 & 0.25 $\pm$ 0.01  & 71.27 $\pm$ 0.07 & 0.84 $\pm$ 0.07 \\
           &            &             &                &      & 1.47 &       & 0.88 $\pm$ 0.01  & 72.48 $\pm$ 0.07 & 0.56 $\pm$ 0.07 \\
           &            &             &                &      & 0.68 &       & 0.58 $\pm$ 0.01  & 73.27 $\pm$ 0.07 & 0.80 $\pm$ 0.07 \\
           & Effelsberg & 49$\arcsec$ & 2022, Jul. 04  & 0.07 & 0.38 & 0.058 & 0.18 $\pm$ 0.01  & 71.22 $\pm$ 0.07 & 0.45 $\pm$ 0.07 \\
           &            &             &                &      & 1.43 &       & 0.89 $\pm$ 0.01  & 72.52 $\pm$ 0.07 & 0.58 $\pm$ 0.07 \\
           &            &             &                &      & 0.67 &       & 0.43 $\pm$ 0.01  & 73.32 $\pm$ 0.07 & 0.60 $\pm$ 0.07 \\
           & Effelsberg & 49$\arcsec$ & 2022, Jul. 11  & 0.07 & 0.28 & 0.064 & 0.25 $\pm$ 0.02  & 71.21 $\pm$ 0.04 & 0.84 $\pm$ 0.08 \\
           &            &             &                &      & 1.04 &       & 0.62 $\pm$ 0.03  & 72.51 $\pm$ 0.01 & 0.56 $\pm$ 0.04 \\
           &            &             &                &      & 0.48 &       & 0.28 $\pm$ 0.03  & 73.23 $\pm$ 0.03 & 0.55 $\pm$ 0.07 \\
           & Effelsberg & 49$\arcsec$ & 2022, Aug. 01  & 0.07 & 0.25 & 0.049 & 0.13 $\pm$ 0.03  & 71.14 $\pm$ 0.03 & 0.50 $\pm$ 0.13 \\
           &            &             &                &      & 0.77 &       & 0.35 $\pm$ 0.02  & 72.50 $\pm$ 0.01 & 0.42 $\pm$ 0.02 \\
           &            &             &                &      & 0.48 &       & 1.00 $\pm$ 0.05  & 72.93 $\pm$ 0.04 & 1.96 $\pm$ 0.10 \\
\hline
\end{tabular}
\tablefoot{The spectral parameters are obtained from Gaussian-fitting.
\tablefoottext{a}{The JVLA spectrum toward Sgr B2(M) is extracted from the Effelsberg beam (FWHM, 49$\arcsec$) sized region.} }
\end{table*}

\begin{table*}[h]
\caption{Summary of NH$_3$ (10,7) maser observations toward Sgr B2(M)}
\centering
\begin{tabular}{lccccccccc}
\hline\hline
Source & Telescope & Beam & Epoch &  Channel  & $S_\nu$  & rms & $\int S_\nu dv$ & $V_{\rm LSR}$ & $\Delta V_{1/2}$  \\
  &   & size & & spacing  &    &  & & &   \\
  &   &  & & (km s$^{-1}$) & (Jy) &  (Jy)  & (Jy km s$^{-1}$) & \multicolumn{2}{c}{ (km s$^{-1}$)}    \\
\hline
\label{spectra107_fitting-sgrb2m}
Sgr B2(M)  & Effelsberg & 49$\arcsec$ & 2021, Feb. 11  & 0.62 & 0.15 & 0.020  & 0.38 $\pm$ 0.06 & 69.77 $\pm$ 0.19 & 2.46 $\pm$ 0.47  \\
           & Effelsberg & 49$\arcsec$ & 2021, Feb. 15  & 0.62 & $\cdots$ & 0.032  & $\cdots$ & $\cdots$ & $\cdots$ \\
           & Effelsberg & 49$\arcsec$ & 2021, Feb. 27  & 0.62 & 0.15 & 0.021  & 0.33 $\pm$ 0.05 & 69.86 $\pm$ 0.14 & 2.09 $\pm$ 0.32 \\
           & Effelsberg & 49$\arcsec$ & 2021, Aug. 12  & 0.07 & 0.14 & 0.041  & 0.08 $\pm$ 0.02 & 69.59 $\pm$ 0.05 & 0.51 $\pm$ 0.17 \\
           &            &             &                &      & 0.14 &        & 0.03 $\pm$ 0.02 & 70.07 $\pm$ 0.03 & 0.18 $\pm$ 0.09 \\
           &            &             &                &      & 0.10 &        & 0.13 $\pm$ 0.04 & 70.80 $\pm$ 0.12 & 1.22 $\pm$ 0.59 \\
  & JVLA\tablefootmark{a}& $0\farcs220~\times~0\farcs094$ & 2022, Mar. 05  & 0.13 & 0.35 & 0.127  & 0.22 $\pm$ 0.06 & 69.55 $\pm$ 0.07 & 0.58 $\pm$ 0.22 \\
           &            &             &                &                          & 0.71 &      & 0.76 $\pm$ 0.06 & 81.90 $\pm$ 0.04 & 1.01 $\pm$ 0.09 \\
           & Effelsberg & 49$\arcsec$ & 2022, Mar. 10  & 0.07 & 0.14 & 0.049 & 0.16 $\pm$ 0.03  & 70.10 $\pm$ 0.08 & 1.08 $\pm$ 0.25 \\
           &            &             &                &      & 0.16 &       & 0.04 $\pm$ 0.01  & 70.80 $\pm$ 0.03 & 0.25 $\pm$ 0.06 \\
           & Effelsberg & 49$\arcsec$ & 2022, Mar. 14  & 0.07 & 0.16 & 0.043 & 0.51 $\pm$ 0.04  & 70.13 $\pm$ 0.11 & 3.02 $\pm$ 0.37 \\
           & Effelsberg & 49$\arcsec$ & 2022, May. 03  & 0.07 & 0.17 & 0.039 & 0.04 $\pm$ 0.02  & 69.26 $\pm$ 0.02 & 0.23 $\pm$ 0.09 \\
           &            &             &                &      & 0.11 &       & 0.07 $\pm$ 0.03  & 69.62 $\pm$ 0.17 & 0.65 $\pm$ 0.18 \\
           &            &             &                &      & 0.15 &       & 0.21 $\pm$ 0.05  & 70.72 $\pm$ 0.11 & 1.28 $\pm$ 0.33 \\
           & Effelsberg & 49$\arcsec$ & 2022, Jun. 17  & 0.07 & 0.20 & 0.051 & 0.15 $\pm$ 0.05  & 69.53 $\pm$ 0.09 & 0.70 $\pm$ 0.22 \\
           &            &             &                &      & 0.16 &       & 0.17 $\pm$ 0.05  & 70.48 $\pm$ 0.16 & 0.99 $\pm$ 0.25 \\
           & Effelsberg & 49$\arcsec$ & 2022, Jul. 04  & 0.07 & 0.17 & 0.063 & 0.22 $\pm$ 0.04  & 69.39 $\pm$ 0.09 & 1.21 $\pm$ 0.30 \\
           & Effelsberg & 49$\arcsec$ & 2022, Jul. 11  & 0.07 & 0.14 & 0.060 & 0.14 $\pm$ 0.03  & 70.30 $\pm$ 0.11 & 0.97 $\pm$ 0.22 \\
           &            &             &                &      & 0.28 &       & 0.07 $\pm$ 0.02  & 70.92 $\pm$ 0.02 & 0.22 $\pm$ 0.06 \\
           & Effelsberg & 49$\arcsec$ & 2022, Aug. 01  & 0.07 & 0.15 & 0.048 & 0.13 $\pm$ 0.01  & 69.54 $\pm$ 0.08 & 0.79 $\pm$ 0.08 \\
           &            &             &                &      & 0.25 &       & 0.14 $\pm$ 0.01  & 70.81 $\pm$ 0.08 & 0.54 $\pm$ 0.08 \\
           &            &             &                &      & 0.07 &       & 0.24 $\pm$ 0.01  & 81.52 $\pm$ 0.08 & 3.15 $\pm$ 0.08 \\
           &            &             &                &      & 0.19 &       & 0.06 $\pm$ 0.01  & 82.10 $\pm$ 0.08 & 0.31 $\pm$ 0.08 \\
\hline
\end{tabular}
\tablefoot{The spectral parameters are obtained from Gaussian-fitting.
\tablefoottext{a}{The JVLA spectrum toward Sgr B2(M) is extracted from the Effelsberg beam (FWHM, 49$\arcsec$) sized region.} }
\end{table*}

\begin{table*}[h]
\caption{1.6 cm JVLA flux densities of individual continuum sources}
\centering
\begin{tabular}{lcccccc}
\hline\hline
\multicolumn{2}{c}{Source} &  R.A. & Dec.  & Deconvolved Size  & P.A.  &    $S_\nu$    \\
 & & ($h\quad m\quad s$) & ($\degr\quad \arcmin\quad \arcsec$) & (arcsec)  & (deg) & (mJy) \\
\hline
\label{continuum_sou}
Sgr B2  & B   & 17 47 19.927 $\pm$ 0.019 & $-$28 23 02.832 $\pm$ 0.041 & (0.796 $\pm$ 0.105) $\times$ (0.331 $\pm$ 0.046) & 15.6  & 181 $\pm$ 22 \\
   & B10.06   & 17 47 19.864 $\pm$ 0.010 & $-$28 23 01.247 $\pm$ 0.013 &       ($\sim$0.22) $\times$ ($\sim$0.12)         &       & 6 $\pm$ 1 \\
        & E   & 17 47 20.116 $\pm$ 0.013 & $-$28 23 08.711 $\pm$ 0.021 & (0.605 $\pm$ 0.056) $\times$ (0.389 $\pm$ 0.036) & 10.5  & 184 $\pm$ 15 \\
        & F1  & 17 47 20.125 $\pm$ 0.053 & $-$28 23 03.920 $\pm$ 0.062 & (1.050 $\pm$ 0.170) $\times$ (0.680 $\pm$ 0.110) & 144.0 & 1090 $\pm$ 170 \\
        & F2  & 17 47 20.167 $\pm$ 0.014 & $-$28 23 03.588 $\pm$ 0.012 & (0.316 $\pm$ 0.047) $\times$ (0.204 $\pm$ 0.060) & 82.0  & 124 $\pm$ 15 \\
        & F3a & 17 47 20.155 $\pm$ 0.012 & $-$28 23 04.878 $\pm$ 0.016 & (0.203 $\pm$ 0.061) $\times$ (0.129 $\pm$ 0.090) & 64.0  & 34 $\pm$ 6 \\
      & F3cd  & 17 47 20.180 $\pm$ 0.014 & $-$28 23 04.628 $\pm$ 0.015 & (0.478 $\pm$ 0.054) $\times$ (0.445 $\pm$ 0.057) & 118.0 & 518 $\pm$ 39 \\
        & F3e & 17 47 20.219 $\pm$ 0.006 & $-$28 23 04.582 $\pm$ 0.013 & (0.206 $\pm$ 0.049) $\times$ (0.145 $\pm$ 0.030) & 1.0   & 32 $\pm$ 4 \\
        & F4  & 17 47 20.218 $\pm$ 0.005 & $-$28 23 04.205 $\pm$ 0.007 & (0.243 $\pm$ 0.029) $\times$ (0.180 $\pm$ 0.032) & 123.0 & 105 $\pm$ 7 \\
     & F10.27 & 17 47 20.081 $\pm$ 0.003 & $-$28 23 05.220 $\pm$ 0.011 &   ($\sim$0.22) $\times$ ($\sim$0.08)                                               &       & 3 $\pm$ 0.6 \\
     & F10.37 & 17 47 20.182 $\pm$ 0.003 & $-$28 23 05.797 $\pm$ 0.005 & (0.151 $\pm$ 0.018) $\times$ (0.050 $\pm$ 0.044) & 103.0 & 30 $\pm$ 2 \\
     & F10.39 & 17 47 20.197 $\pm$ 0.001 & $-$28 23 06.484 $\pm$ 0.004 &       ($\sim$0.078) $\times$ ($\sim$0.055)       &       & 12 $\pm$ 1 \\
        & G   & 17 47 20.290 $\pm$ 0.014 & $-$28 23 02.936 $\pm$ 0.015 & (0.333 $\pm$ 0.055) $\times$ (0.179 $\pm$ 0.067) & 55.0  & 93 $\pm$ 13 \\
        & I   & 17 47 20.385 $\pm$ 0.064 & $-$28 23 05.167 $\pm$ 0.071 & (3.88 $\pm$ 0.20) $\times$ (1.293 $\pm$ 0.10)    & 40.1  & 3740 $\pm$ 190 \\
     & I10.52 & 17 47 20.332 $\pm$ 0.006 & $-$28 23 08.006 $\pm$ 0.013 &       ($\sim$0.140) $\times$ ($\sim$0.093)       &       & 8 $\pm$ 2 \\
        & K1  & 17 47 19.781 $\pm$ 0.009 & $-$28 22 20.588 $\pm$ 0.022 & (0.612 $\pm$ 0.056) $\times$ (0.285 $\pm$ 0.026) & 8.9   & 123 $\pm$ 10 \\
        & K2  & 17 47 19.883 $\pm$ 0.002 & $-$28 22 18.412 $\pm$ 0.007 & (0.424 $\pm$ 0.018) $\times$ (0.088 $\pm$ 0.003) & 164.1 & 32 $\pm$ 1 \\
        & K3  & 17 47 19.903 $\pm$ 0.013 & $-$28 22 17.051 $\pm$ 0.014 & (0.442 $\pm$ 0.054) $\times$ (0.429 $\pm$ 0.056) & 129.0 & 181 $\pm$ 13 \\
        & K4  & 17 47 20.017 $\pm$ 0.015 & $-$28 22 04.474 $\pm$ 0.020 & (1.196 $\pm$ 0.049) $\times$ (0.848 $\pm$ 0.034) & 159.8 & 289 $\pm$ 12 \\
        & K7  & 17 47 19.905 $\pm$ 0.008 & $-$28 22 13.401 $\pm$ 0.027 & (0.347 $\pm$ 0.081) $\times$ (0.104 $\pm$ 0.018) & 166.2 & 7 $\pm$ 1 \\
    & Z10.24  & 17 47 20.043 $\pm$ 0.001 & $-$28 22 41.143 $\pm$ 0.003 & (0.175 $\pm$ 0.013) $\times$ (0.111 $\pm$ 0.007) & 147.9 & 23 $\pm$ 1 \\
        & S   & 17 47 20.472 $\pm$ 0.041 & $-$28 23 45.120 $\pm$ 0.034 & (1.747 $\pm$ 0.113) $\times$ (0.840 $\pm$ 0.057) & 54.0  & 1231 $\pm$ 79 \\
\hline
\end{tabular}
\end{table*}

\begin{table*}[h]
\caption{NH$_3$ maser positions in Sgr B2, derived from the JVLA observations.}
\centering
\begin{tabular}{cccccccc}
\hline\hline
Transition &  &  R.A. & Dec. &  $S_\nu$  & $T_{\rm MB}$  & $V_{\rm LSR}$ & $\Delta V_{1/2}$  \\
 & &  & &    &    &   &   \\
 &  & ($h\quad m\quad s$) & ($\degr\quad \arcmin\quad \arcsec$) & (Jy~beam$^{-1}$) & (10$^4$ K) &   \multicolumn{2}{c}{ (km~s$^{-1}$)}    \\
\hline
\label{NH3-positions}
(6,3)      & 63A\tablefootmark{+}  & 17 47 20.005 $\pm$ 0.003 & $-$28 22 17.285 $\pm$ 0.009 & 0.058 & 1.026  & 62.09 $\pm$ 0.02 & 0.60 $\pm$ 0.04 \\
           & 63B\tablefootmark{+}  & 17 47 19.918 $\pm$ 0.002 & $-$28 22 17.037 $\pm$ 0.008 & 0.085 & 1.503  & 68.94 $\pm$ 0.01 & 0.40 $\pm$ 0.02 \\
           & 63C\tablefootmark{x}  & 17 47 20.036 $\pm$ 0.001 & $-$28 22 41.135 $\pm$ 0.004 & 0.114 & 2.028  & 57.26 $\pm$ 0.01 & 0.49 $\pm$ 0.01 \\
           & 63D\tablefootmark{*}  & 17 47 20.200 $\pm$ 0.002 & $-$28 23 06.500 $\pm$ 0.008 & 0.019 & 0.345   & 69.02 $\pm$ 0.12 & 1.14 $\pm$ 0.12 \\
           &     &                    &                             & 0.048 & 0.850   & 70.22 $\pm$ 0.12 & 0.62 $\pm$ 0.12 \\
           &     &                    &                             & 0.027 & 0.470   & 71.00 $\pm$ 0.12 & 0.48 $\pm$ 0.12 \\
           &     &                    &                             & 0.098 & 1.720  & 71.80 $\pm$ 0.12 & 0.44 $\pm$ 0.12 \\
           &     &                    &                             & 0.102 & 1.774  & 72.68 $\pm$ 0.12 & 0.75 $\pm$ 0.12 \\
(7,4)      & 74A\tablefootmark{+}  & 17 47 20.005 $\pm$ 0.001 & $-$28 22 17.255 $\pm$ 0.002 & 0.381 & 6.771 & 62.58 $\pm$ 0.002 & 0.37 $\pm$ 0.005 \\
           & 74B\tablefootmark{*} & 17 47 20.200 $\pm$ 0.001 & $-$28 23 06.512 $\pm$ 0.005 & 0.125 & 2.219 & 72.55 $\pm$ 0.006 & 0.72 $\pm$ 0.02 \\
           & 74C\tablefootmark{**} & 17 47 20.471 $\pm$ 0.001 & $-$28 23 45.783 $\pm$ 0.005 & 0.177 & 3.144 & 77.41 $\pm$ 0.004 & 0.54 $\pm$ 0.01 \\
(8,5)      & 85A\tablefootmark{+}   & 17 47 19.862 $\pm$ 0.003 & $-$28 22 18.254 $\pm$ 0.011 & 0.019 & 0.340   & 54.32 $\pm$ 0.06 & 0.88 $\pm$ 0.17  \\
           &     &                         &                             & 0.019 & 0.350   & 55.14 $\pm$ 0.04 & 0.34 $\pm$ 0.07 \\
           &     &                         &                             & 0.041 & 0.742   & 56.52 $\pm$ 0.04 & 2.12 $\pm$ 0.14 \\
           & 85B\tablefootmark{*} & 17 47 20.118 $\pm$ 0.005 & $-$28 23 06.230 $\pm$ 0.015 & 0.024 & 0.444   & 70.95 $\pm$ 0.04 & 0.95 $\pm$ 0.08  \\
(9,6)      & 96A\tablefootmark{+}  & 17 47 19.876 $\pm$ 0.002 & $-$28 22 17.903 $\pm$ 0.009 & 0.070 & 1.207  & 77.92 $\pm$ 0.02 & 0.94 $\pm$ 0.05 \\
           & 96B\tablefootmark{+}   & 17 47 19.855 $\pm$ 0.003 & $-$28 22 17.977 $\pm$ 0.010 & 0.083 & 1.416  & 63.78 $\pm$ 0.01 & 0.82 $\pm$ 0.03 \\
           & 96C\tablefootmark{*} & 17 47 20.200 $\pm$ 0.001 & $-$28 23 06.511 $\pm$ 0.005 & 0.217 & 3.712  & 71.24 $\pm$ 0.13 & 0.73 $\pm$ 0.13 \\
           &     &                          &                             & 3.211 & 55.15 & 72.48 $\pm$ 0.13 & 0.56 $\pm$ 0.13 \\
           &     &                          &                             & 1.208 & 20.59 & 73.24 $\pm$ 0.13 & 0.59 $\pm$ 0.13 \\
           & 96D\tablefootmark{**} & 17 47 20.471 $\pm$ 0.002 & $-$28 23 45.763 $\pm$ 0.008 & 0.172 & 3.214  & 77.38 $\pm$ 0.01 & 0.66 $\pm$ 0.03 \\
           &     &                         &                             & 0.066 & 1.097  & 78.20 $\pm$ 0.09 & 1.27 $\pm$ 0.14 \\
           &     &                         &                             & 0.079 & 1.378  & 82.60 $\pm$ 0.01 & 0.81 $\pm$ 0.03 \\
           &     &                         &                             & 0.064 & 1.086  & 83.52 $\pm$ 0.01 & 0.90 $\pm$ 0.06 \\
(10,7)     & 107A\tablefootmark{+}   & 17 47 19.859 $\pm$ 0.002 & $-$28 22 12.925 $\pm$ 0.010 & 0.069 & 1.257  & 65.55 $\pm$ 0.02  & 0.57 $\pm$ 0.04 \\
           & 107B\tablefootmark{+}   & 17 47 19.259 $\pm$ 0.001 & $-$28 22 14.724 $\pm$ 0.003 & 3.710 & 67.17 & 81.98 $\pm$ 0.001 & 0.96 $\pm$ 0.002 \\
           &    &                          &                               & 0.263 & 4.770  & 86.35 $\pm$ 0.01  & 0.60 $\pm$ 0.02 \\
           & 107C\tablefootmark{+}   & 17 47 19.862 $\pm$ 0.003 & $-$28 22 18.257 $\pm$ 0.011 & 0.041 & 0.751   & 55.08 $\pm$ 0.03  & 1.39 $\pm$ 0.10 \\
           &    &                            &                               & 0.029 & 0.520   & 56.36 $\pm$ 0.03  & 0.46 $\pm$ 0.07 \\
           &    &                            &                               & 0.053 & 0.961   & 57.24 $\pm$ 0.03  & 1.21 $\pm$ 0.09 \\
           & 107D\tablefootmark{*} & 17 47 20.117 $\pm$ 0.002 & $-$28 23 06.216 $\pm$ 0.006 & 0.165 & 2.967  & 69.48 $\pm$ 0.13  & 0.72 $\pm$ 0.13 \\
           &    &                        &                               & 0.139 & 2.706  & 70.23 $\pm$ 0.13  & 0.61 $\pm$ 0.13 \\
           &    &                        &                               & 0.137 & 2.446  & 70.89 $\pm$ 0.13  & 0.69 $\pm$ 0.13 \\
           & 107E\tablefootmark{**} &17 47 20.471 $\pm$ 0.002 & $-$28 23 45.779 $\pm$ 0.007 & 0.123 & 2.224  & 78.42 $\pm$ 0.01  & 1.25 $\pm$ 0.02 \\
\hline
\end{tabular}
\tablefoot{\tablefoottext{+}{These nine maser spots originate in Sgr B2(N).} \tablefoottext{x}{63C arises in Sgr B2(NS).} \tablefoottext{*}{These five maser sources belong to Sgr B2(M).} \tablefoottext{**}{These three maser spots originate in Sgr B2(S).} The flux density scale calibration accuracy is estimated to be within 15\%.
}
\end{table*}

\end{appendix}

\end{document}